\documentclass[10pt]{article}
\usepackage[utf8]{inputenc}
\usepackage[left=1in, right=1in, top=1in, bottom=1in]{geometry}
\usepackage{url} 
\usepackage{amsfonts}   
\usepackage{amsmath} 
\usepackage{amsthm} 
\usepackage{mathrsfs} 
\usepackage{graphicx} 
\usepackage{natbib}
\usepackage{tcolorbox}

\newcommand{\bx}{\boldsymbol x}
\newcommand{\bv}{\boldsymbol v}

\newcommand{\bb}{\boldsymbol b}
\newcommand{\bw}{\boldsymbol w}

\title{Adaptive Gaussian Markov Random Fields for Child Mortality Estimation}

\author{Serge Aleshin-Guendel$^{1}$\thanks{To whom correspondence should be addressed. This report is released to inform interested parties of ongoing research and to encourage discussion of work in progress. The views expressed are those of the authors and not those of the U.S. Census Bureau.}, and Jon Wakefield$^{2, 3}$ \\ 
$^{1}$Center for Statistical Research and Methodology, U.S. Census Bureau,\\
Washington, D.C., U.S.A. \\
$^{2}$Department of Biostatistics, University of Washington, Seattle, Washington, U.S.A. \\ $^{3}$Department of Statistics, University of Washington, Seattle, Washington, U.S.A.}

\begin{document}
\maketitle

\begin{abstract}
{
The under-5 mortality rate (U5MR), a critical health indicator, is typically estimated from household surveys in lower and middle income countries. Spatio-temporal disaggregation of household survey data can lead to highly variable estimates of U5MR, necessitating the usage of smoothing models which borrow information across space and time. The assumptions of common smoothing models may be unrealistic when certain time periods or regions are expected to have shocks in mortality relative to their neighbors, which can lead to oversmoothing of U5MR estimates. In this paper, we develop a spatial and temporal smoothing approach based on Gaussian Markov random field models which incorporate knowledge of these expected shocks in mortality. We demonstrate the potential for these models to improve upon alternatives not incorporating knowledge of expected shocks in a simulation study. We apply these models to estimate U5MR in Rwanda at the national level from 1985-2019, a time period which includes the Rwandan civil war and genocide.
} 
\end{abstract}

\section{Introduction}
The under-5 mortality rate (U5MR) is an important indicator for summarizing the health of a country. This is highlighted by the United Nations (UN) Sustainable Development Goals (SDGs) in SDG 3.2, which states ``By 2030, end preventable deaths of newborns and children under 5 years of age, with all countries aiming to reduce neonatal mortality to at least as low as 12 per 1,000 live births and under-5 mortality to at least as low as 25 per 1,000 live births" \citep{UN_2015}. 
Due to a lack of vital registration systems in many lower and middle income countries, U5MR is typically estimated from household surveys, like the Demographic and Health Surveys (DHS). However, the reliable estimation of U5MR from such household surveys at fine spatio-temporal scales require the usage of smoothing models which borrow information across space and time. For example, even at the temporal scale, the United Nations Inter-Agency Group for Child Mortality Estimation (UN IGME) typically aggregates survey data to five year periods in order to produce estimates with acceptable precision \citep{Pedersen_2012, UN_2019}.

The statistical methods for estimating U5MR over space and/or time typically accomplish spatio-temporal smoothing through the use of random effects based on Gaussian Markov random fields (GMRFs) or closely related models. In particular, UN IGME produces national level estimates of U5MR yearly using the Bayesian B-spline bias-reduction (B3) method \citep{Alkema_2014, Alkema_2014a}, which uses smoothing splines that have well-known connections to GMRFs \citep[see e.g.][]{Yue_2014}, and has supported and produced subnational estimates of U5MR using GMRFs \citep{Li_2019, UN_2021}. The assumptions underlying these GMRF-based smoothing models may not be realistic for estimating U5MR when certain time periods or regions are expected to have shocks (i.e. large changes) in mortality relative to their neighbors. 
In this paper, we are motivated by U5MR estimation in two contexts where such shocks in mortality could occur: 1) Rwanda, where a civil war and genocide took place in the mid 1990's, and 2) multi-country models, which have become more common in the global health literature \citep{Burstein_2019}, where we simultaneously estimate U5MR subnationally across multiple countries.
In such scenarios, GMRF-based smoothing models may lead to oversmoothing of U5MR estimates in certain time periods or regions.


In this article, we develop smoothing models which incorporate knowledge of expected shocks in mortality, but still allow information to be borrowed across space and time. We discuss our motivating applications and review GMRFs and U5MR estimation in Section \ref{sec:background}. In Section \ref{sec:agmrfs}, we extend commonly used GMRFs to allow the incorporation of knowledge of expected shocks in mortality, which we call adaptive  Gaussian  Markov  random  fields (AGMRFs). Section \ref{sec:misc} provides details of implementing our AGMRFs. Section \ref{sec:simulations} presents a simulation study assessing how AGMRFs can improve the performance of a model used to estimate U5MR. Finally, we apply our AGMRFs to estimate U5MR in Rwanda in Section
\ref{sec:rwanda_app} and in a multicountry setting in Appendix E of the Supplementary Materials.

\section{Background and Motivating Applications}
\label{sec:background}

In this section we first review GMRFs and U5MR estimation using the smoothed direct model of \cite{Mercer_2015}, and then provide two motivating applications involving U5MR.

\subsection{Gaussian Markov Random Fields}
Statistical models for child mortality estimation typically involve random effects in space and/or time. This is the case for the smoothed direct model of \cite{Mercer_2015}, which we review in the following section. In most cases, these random effects are GMRFs, Gaussian random vectors defined on labelled graphs where sparsity in the precision matrix implies certain conditional independence properties \citep{Rue_2005}. Typically, these spatial and temporal random effects are improper GMRFs. In this section, we review the first-order random walk temporal random effect and the intrinsic conditional autoregression spatial random effect. For a full review of GMRFs, improper GMRFs, and their various properties, we refer the reader to Chapters 2 and 3 of \cite{Rue_2005}. 

\subsubsection{First-Order Random Walk}
\label{sec:rw1}
Suppose we have $N$ time periods and we specify a structured temporal random effect $\bx=(x_1,\dots,x_N)$. Suppose for $i=1,\dots, N-1$ we specify 
\[x_{i+1}\mid x_i, \tau \sim\textsf{Normal}(x_i,\tau^{-1}),
\]
where $\tau$ is the precision for each transition. This is referred to as a first-order random walk or random walk 1 (RW1), which we denote as $\bx\sim \textsf{RW1}(\tau)$ \citep[see e.g. page 95 of ][]{Rue_2005}. This leads to a density for $\bx$ of
\[
p(\bx\mid \tau)\propto \tau^{(N-1)/2}\exp\left\{-\frac{\tau}{2}\sum_{i=1}^{N-1}(x_{i+1}-x_i)^2\right\}= \tau^{(N-1)/2}\exp\left(-\frac{1}{2}\bx^{T}Q\bx\right),
\]
where $Q=\tau R$ is a precision matrix determined by the structure matrix $R$ with
\[
R_{ij}=
    \begin{cases}
      1, & \text{if}\ i=j=1\ \text{or}\ i=j=N, \\
      2, & \text{if}\ i=j=k\ \text{and}\ k\notin\{1,N\},\\
      -1, & \text{if}\ j=i+1\ \text{or}\ i=j+1,\\
      0, & \text{if}\ |i-j|>1.
    \end{cases}
\]
It can be verified that $Q\boldsymbol{1}=\boldsymbol{0}$, and so $Q$ is rank $N-1$ and RW1s are improper GMRFs. It follows that $p(\bx\mid \tau)$ is invariant to an addition of a constant vector to $\bx$, thus when a RW1 is included in a model with an intercept we enforce the sum-to-zero constraint $\sum_{i=1}^Nx_i=0$.

\subsubsection{Intrinsic Conditional Autoregression}
\label{sec:icar}
Suppose we have $N$ areal units and we specify a structured spatial random effect $\bx=(x_1,\dots,x_N)$. We assume there are no islands, i.e. we assume the graph of the areal units is connected. See \cite{Freni_2018} for more details when there are islands. Suppose for neighboring regions $i\sim j$ we specify
\[x_i-x_j\mid \tau \sim\textsf{Normal}(0,\tau^{-1}),
\]
where $\tau$ is the precision for each difference. This is referred to as an intrinsic conditional autoregression (ICAR), which we denote as $\bx\sim \textsf{ICAR}(\tau)$ \citep[see e.g. page 101 of ][]{Rue_2005}. This leads to a density for $\bx$ of
\[
p(\bx\mid \tau)\propto \tau^{(N-1)/2} \exp\left(-\frac{1}{2}\bx^{T}Q\bx\right),
\]
where $Q=\tau R$ is a precision matrix determined by the structure matrix $R$ with
\[
R_{ij}=
    \begin{cases}
     n_i, & \text{if}\ i=j, \\
      -1, & \text{if}\ i\sim j,\\
      0, & \text{otherwise},
    \end{cases}
\]
where $n_i$ is the number of regions neighbouring $i$.
It can be verified that $Q\boldsymbol{1}=\boldsymbol{0}$, and so $Q$ is rank $N-1$ and ICARs are improper GMRFs. It follows that $p(\bx\mid \tau)$ is invariant to an addition of a constant vector to $\bx$, thus when a ICAR is included in a model with an intercept we enforce the sum-to-zero constraint $\sum_{i=1}^Nx_i=0$. As time periods can be viewed as areal units with a specific neighborhood structure, RW1s are special cases of ICARs.

\subsection{The Smoothed Direct Model}
\label{sec:smoothed_direct}
In this section, we review the smoothed direct model of \cite{Mercer_2015} for child mortality estimation, a recent extension of the seminal Fay--Herriot model \citep{Fay_1979}. Suppose for a country of interest we have $S$ different surveys of full birth histories, each of which can be used to produce, in the language of small area estimation \citep{Rao_2015}, direct estimates of U5MR, i.e. survey weighted estimates of U5MR with associated design-based standard errors. For ease of exposition, we focus on estimating U5MR either at the national level over multiple time periods, or at the subnational level in one time period. While one could use these direct estimates as estimates of U5MR, the temporal or spatial disaggregation of the data can lead to noisy estimates with large standard errors. In the most extreme case, there can be no data in a particular time period or region. It is thus desirable to borrow information across time or space to smooth these estimates.

Let $N$ denote either the number of time periods in a national level model or the number of administrative regions in a subnational level model, which we refer to generically as areas. For $s\in\{1,\dots,S\}$ and $i\in\{1,\dots,N\}$, let $\hat{p}_{is}$ denote the direct estimate of U5MR in area $i$ from survey $s$. Let $y_{is}=\text{logit}\left(\hat{p}_{is}\right)$, and let $\hat{V}_{is}$ denote the design-based variance associated with $y_{is}$. 
\cite{Mercer_2015} used the asymptotic distribution of $y_{is}$ as a working likelihood:
\[
y_{is}\mid \eta_{is}\sim\textsf{Normal}(\eta_{is},\hat{V}_{is}),
\]
where $\hat{V}_{is}$ is treated as fixed and known.
Smoothing of the direct estimates over the areas $i$ is accomplished through a model for $\eta_{is}$,
\begin{equation*}
    \eta_{is} =\mu + v_i + x_i + \nu_s,
\end{equation*}
where $\nu_s \stackrel{iid}{\sim}\textsf{Normal}(0,\sigma^2_{\nu})$ is a survey random effect (if $S$ is small this can be replaced by a fixed effect with a sum-to-zero constraint), $v_i\stackrel{iid}{\sim}\textsf{Normal}(0,\sigma^2_{v})$ is an unstructured area-level random effect, and $\bx$ is a structured area-level random effect, either $\bx\sim \textsf{RW1}(\tau_{x})$ if we are working with the national level model or $\bx\sim \textsf{ICAR}(\tau_{x})$  if we are working with the subnational level model.

A Bayesian approach is taken, where priors are set on the intercept $\mu$ and all precision parameters. Smoothed estimates of U5MR in area $i$ are then based on the posterior $p(p_{i}\mid \{y_{is}\}_{is})$, where 
\[p_{i}=\text{expit}(\mu + v_i + x_i). \]

The total area-level random effect in the smoothed direct model, $\bb = \bv + \bx$, which is the sum of an unstructured random effect and a structured random effect, is due to Besag, York,  and Molli{\'e} (BYM) \citep{Besag_1991}. We reparameterize the total area-level random effect, $\bb$, as a so-called BYM2, following \cite{Riebler_2016} and \cite{Simpson_2017}. In this parameterization, rather than representing the total area-level random effect $\bb$ as a a sum of an unstructured random effect, $\bv$, and a structured random effect, $\bx$, with independent precision parameters, we represent the total area-level random effect as
$\bb = (\sqrt{1-\phi}\bv + \sqrt{\phi}\bx^*)/\sqrt{\tau_b}$, where  $v_i\stackrel{iid}{\sim}\textsf{Normal}(0,1)$ and $\bx^*\sim\textsf{Normal}(0, R_{\star}^-)$. Here $R_{\star}$ is the structure matrix of the RW1 or ICAR scaled as in \cite{Sorbye_2014}, and $R_{\star}^-$ is the generalized inverse of $R_{\star}$. $1/\tau_b$ can then be interpreted as the marginal variance of $\bb$, and $\phi$ can be interpreted as the fraction of the variance explained by the structured random effect. Penalized complexity priors \citep{Simpson_2017} are then adopted for $\tau_b$ and $\phi$.

The smoothed direct model is a latent Gaussian model, i.e. the data, $y_{is}$, are conditionally independent given a latent Gaussian vector, $\{\eta_{is}\}_{is}$, with a small number of hyperparameters. Thus, the smoothed direct model can be fit using the Integrated Nested Laplace Approximation (INLA) \citep{Rue_2009}, an alternative to Markov chain Monte Carlo (MCMC) methods for fitting Bayesian hierarchical models, using the $\texttt{R}$ package $\texttt{R-INLA}$. In the case of latent Gaussian models, INLA is typically as fast or faster than MCMC for model fitting, while also being essentially fully automated \citep[see e.g.,][]{Fong_2010}. It can be difficult to fully automate MCMC due to the need to diagnose convergence. Fast and automated computation is important in the context of child mortality estimation, as it would be ideal for individuals in settings with minimal computational resources to be able to fit smoothing models. The \texttt{R} package \texttt{SUMMER} \citep{Li_2020} contains implementations of the smoothed direct model and has been used to produce estimates of U5MR supported by the UN IGME \citep{Li_2019}.

\subsection{Child Mortality Estimation: Two Motivating Examples}
\label{sec:motivation}
In this section we provide two motivating child mortality applications where the assumptions underlying the RW1 and ICAR random affects used in the smoothed direct model may not be realistic.

\subsubsection{A National Model for Rwanda}
\label{sec:rwanda_ex}
Suppose we would like yearly estimates of U5MR at the national level in Rwanda from 1985 through 2015, the last year with a DHS survey, and predictions of U5MR from 2016 through 2019. Rwanda experienced a civil war from 1990-1994, culminating in the Rwandan genocide in 1994 \citep{De_2010, Nyseth_2017}. The civil war, the genoicde, and their aftermath produced a shock in child mortality in Rwanda in the 90s.

\begin{figure}[!ht]
\centering
\includegraphics[width=0.75\linewidth]{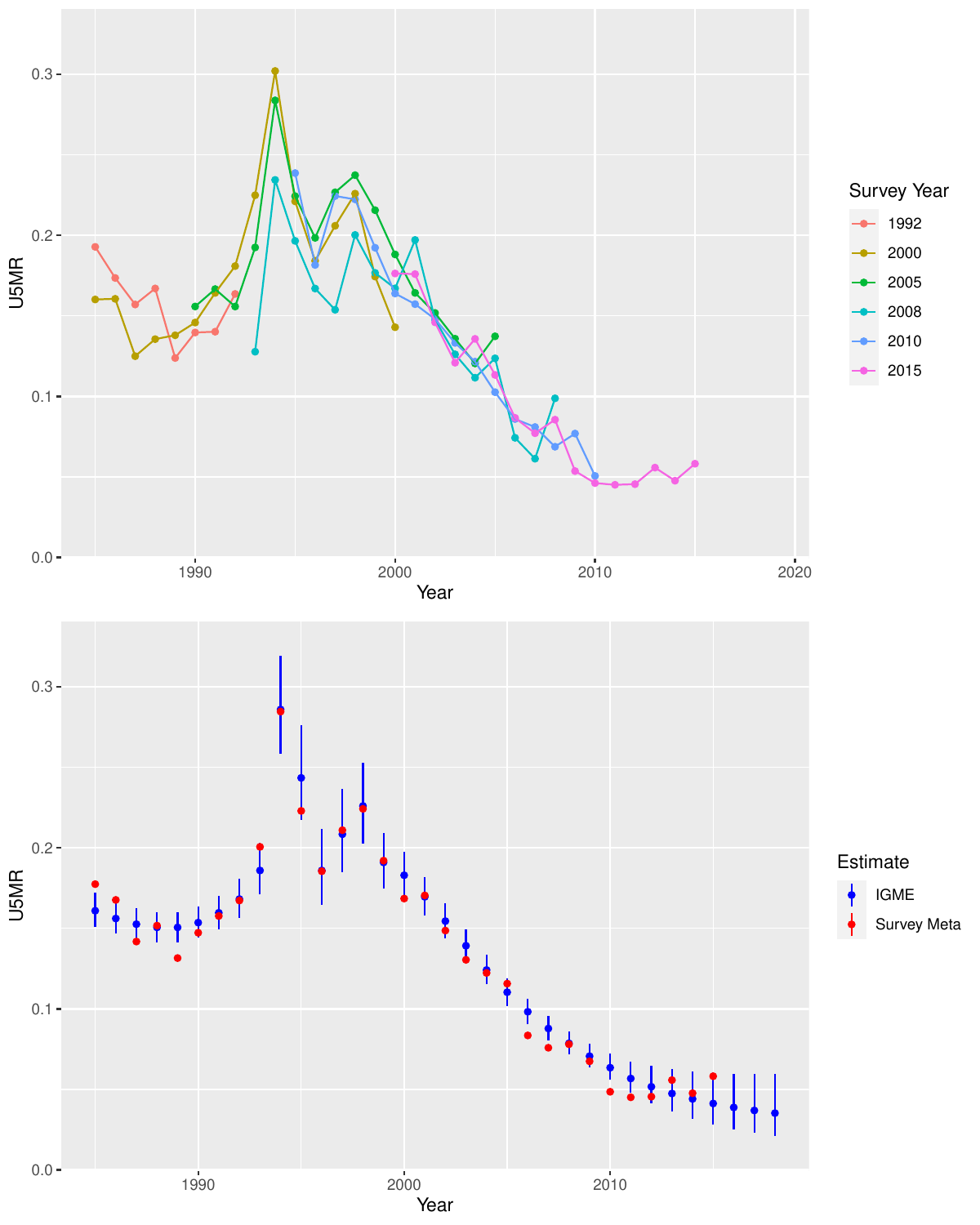}
\caption{Top Panel: Direct estimates of U5MR for Rwanda from six DHS surveys, for the 15 years prior to each survey. Bottom Panel: UN IGME's U5MR estimates and 90\% uncertainty intervals, and a meta-analysis estimator of U5MR based on the direct estimates in the top panel. }
\label{fig:rwanda_direct}
\end{figure}

In the DHS, women between the ages of 15 and 49 are asked their full birth history, so that the available data are birth dates and (if applicable) death dates. This retrospective nature allows one to reconstruct mortality estimates backwards in time.
In Figure \ref{fig:rwanda_direct}, we plot the direct estimates of U5MR for Rwanda from the six DHS surveys between 1992 and 2015 (DHS typically carries out surveys every five years), for the 15 years prior to each survey. The shock to child mortality in the mid 90s is clear from this plot. Further, we plot UN IGME's 2019 estimates of U5MR, which are based on the B3 model, and overlay a ``meta-analysis" estimator of U5MR based on the direct estimates from the six DHS surveys, i.e., the precision-weighted average of each survey's estimate in each year. We note here that the B3 model uses an ad-hoc approach to produce estimates for what it deems to be conflict years, which are 1993-1999 for Rwanda \citep{UN_2019}. Rather than fitting a model to all of the available data, it fits a model leaving out data from conflict years. From the model, it predicts U5MR for the conflict years, and then adds on a separate conflict-specific mortality rate to the estimate for conflict years. This conflict-specific mortality rate may be based on the data used to fit the rest of the model, or based on other outside sources, and does not include any uncertainty.

We would like to modify the national level smoothed direct model so that it does not oversmooth U5MR during the years in which we expect shocks in mortality, which we refer to as conflict years. In more detail, when using a RW1 prior for the structured temporal effect in the smoothed direct model, we do not believe it is likely that transitions not involving conflict years will have the same variance as transitions involving conflict years. We expect transitions not involving conflict years to have a smaller variance than transitions involving conflict years. In other words, the transitions of the structured temporal effect should not be exchangeable as is assumed in the RW1.

\subsubsection{A Multi-country Subnational Model}
\label{sec:multicountry_ex}
\begin{figure}[!ht]
\centering
\includegraphics[width=0.95\linewidth]{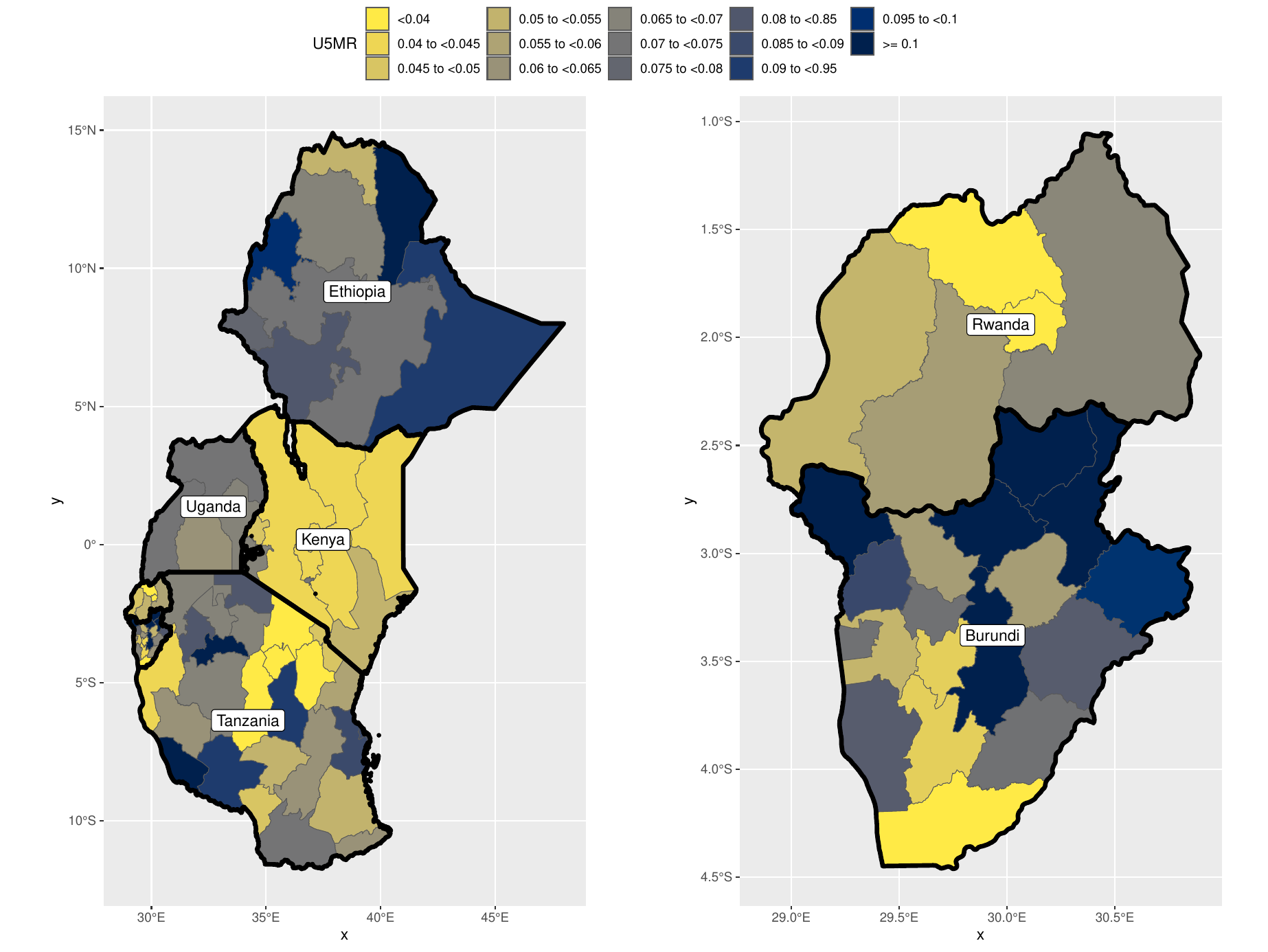}
\caption{Left Panel: Direct estimates of U5MR for the 2010-2014 time period for Burundi, Ethiopia, Kenya, Rwanda, Tanzania, and Uganda. Right Panel: Zoomed in direct estimates of U5MR for Burundi and Rwanda.}
\label{fig:mc_direct}
\end{figure}

Suppose we would like to simultaneously estimate U5MR subnationally at the Admin1 level across multiple countries during the 2010-2014 time period. Specifically, we consider Burundi, Ethiopia, Kenya, Rwanda, Tanzania, and Uganda using DHS surveys from 2016, 2016, 2014, 2015, 2015, and 2016 respectively. Due to the country boundaries, Admin1 regions within the same country are likely to have more similar outcomes than Admin1 regions not within the same country. In Figure \ref{fig:mc_direct}, we plot the direct estimates of U5MR at the Admin1 level for Burundi, Ethiopia, Kenya, Rwanda, Tanzania, and Uganda. We see that along some borders, the Admin1 U5MR direct estimates are fairly similar (e.g. Uganda and Tanzania), and along other borders there's a large difference (e.g. Kenya and Ethiopia).

We would like to modify the subnational level smoothed direct model so that it does not oversmooth U5MR over country boundaries. In more detail, when using an ICAR prior for the structured spatial effect in the smoothed direct model, we do not believe it is likely that differences involving neighboring Admin1 regions within the same country will have the same variance as differences involving neighboring Admin1 regions not within the same country. We expect differences involving neighboring Admin1 regions within the same country to have a smaller variance than differences involving neighboring Admin1 regions not within the same country. In other words, the differences of the structured temporal effect should not be exchangeable as is assumed in the ICAR.

\section{Adaptive Gaussian Markov Random Fields}
\label{sec:agmrfs}
Motivated by the child mortality examples described in the previous section, in this section we develop what we call adaptive Gaussian Markov random fields (AGMRFs). We first develop general first order AGMRFs, before focusing on two specific use cases for the Rwandan and multi-country child mortality applications. We conclude by discussing connections to previous work. In Appendix A of the Supplementary Materials, we consider possible extensions to higher order AGMRFs.

\subsection{General First Order Adaptive Gaussian Markov Random Fields} \label{sec:general_agmrfs}

\subsubsection{Adaptive First-Order Random Walk}
Suppose we have $N$ time periods and we specify a structured temporal random effect $\bx=(x_1,\dots,x_N)$. Suppose for $i=1,\dots, N-1$ we specify 
\begin{equation} \label{eq:arw1}
    x_{i+1}\mid x_i, \tau_i \sim\textsf{Normal}(x_i,\tau_i^{-1}),
\end{equation}
where $\tau_i$ is the precision when moving from time $i$ to time $i+1$.  We refer to this as an adaptive first-order random walk or an adaptive random walk 1 (ARW1). This leads to a density for $\bx$ of
\[
p(\bx\mid \tau_1,\dots, \tau_{N-1})\propto \left(\prod_{i=1}^{N-1}\tau_i^{1/2}\right)\exp\left\{-\frac{1}{2}\sum_{i=1}^{N-1}\tau_i(x_{i+1}-x_i)^2\right\}= \left(\prod_{i=1}^{N-1}\tau_i^{1/2}\right)\exp\left(-\frac{1}{2}\bx^{T}Q\bx\right),
\]
where $Q$ is a precision matrix with the same sparsity structure as the precision matrix of a RW1, such that 
\[
Q_{ij}=
    \begin{cases}
      \tau_1, & \text{if}\ i=j=1, \\
      \tau_{N-1}, & \text{if}\ i=j=N, \\
      \tau_{i-1}+\tau_{i}, & \text{if}\ i=j=k\ \text{and}\ k\notin\{1,N\},\\
      -\tau_i, & \text{if}\ j=i+1\ \text{or}\ i=j+1,\\
      0, & \text{if}\ |i-j|>1.
    \end{cases}
\]
It can be verified that $Q\boldsymbol{1}=\boldsymbol{0}$, and so $Q$ is rank $N-1$ and ARW1s are improper GMRFs. It follows that $p(\bx\mid \tau)$ is invariant to an addition of a constant vector to $\bx$, thus when an ARW1 is included in a model with an intercept, we enforce the sum-to-zero constraint $\sum_{i=1}^Nx_i=0$.

\subsubsection{Adaptive Intrinsic Conditional Autoregression}
Suppose we have $N$ areal units with no islands and we specify a spatial random effect $\bx=(x_1,\dots,x_N)$. Suppose, for neighboring regions $i\sim j$, we specify
\begin{equation}
    \label{eq:aicar}
x_i-x_j\mid \tau_{ij} \sim\textsf{Normal}(0,\tau_{ij}^{-1}),
\end{equation}
where $\tau_{ij}$ is the precision for the difference between units $i$ and $j$. We refer to this as an adaptive intrinsic conditional autoregression (AICAR). This leads to a density for $\bx$ of
\[
p(\bx\mid \{\tau_{ij}\}_{i\sim j})\propto (|Q|^*)^{1/2} \exp\left(-\frac{1}{2}\bx^{T}Q\bx\right),
\]
where $|\cdot|^*$ denotes the generalized determinant of a matrix \citep[see Chapter 3 of ][]{Rue_2005}, defined as the product of its non-zero eigenvalues, and
$Q$ is a precision matrix with the same sparsity structure as the precision matrix of an ICAR, such that
\[
Q_{ij}=
    \begin{cases}
      \sum_{k\mid i\sim k}\tau_{ik}, & \text{if}\ i=j, \\
      -\tau_{ij}, & \text{if}\ i\sim j,\\
      0, & \text{otherwise}.
    \end{cases}
\]
It can be verified that $Q\boldsymbol{1}=\boldsymbol{0}$, and so $Q$ is rank $N-1$ and AICARs are improper GMRFs. It follows that $p(\bx\mid \tau)$ is invariant to an addition of a constant vector to $\bx$, thus when an AICAR is included in a model with an intercept, we enforce the sum-to-zero constraint $\sum_{i=1}^Nx_i=0$. 

\subsection{Two Specific First Order Adaptive Gaussian Markov Random Fields}
\label{sec:usecases}
The general AGMRFs introduced in the previous section have a large number of precision parameters, which makes them very flexible. However, this flexibility can cause difficulties for prior specification and computation. It is difficult in practice to specify substantive priors for a large number of hyperparameters. In addition, one needs a small number of hyperparameters if INLA is to be used for computation. In this section, we specialize these general AGMRFs to our motivating child mortality applications from Section \ref{sec:motivation}, drastically reducing the number of precision parameters.  This specialization sacrifices flexibility in our model so that it is parsimonious, more amenable to substantive prior specification, and able to be fit in INLA.

\subsubsection{An Adaptive RW1 for Conflicts} \label{sec:usecase_conflict}
Suppose we are specifying a structured temporal random effect for $N$ years, where some subset of the $N$ years are conflict years. In the case of Rwanda, the UN IGME categorizes 1993-1999 as conflict years. Let $C$ denote the subset of conflict years. Suppose for $i=1,\dots, N-1$ we specify
\begin{align*}
    x_{i+1}\mid x_i, \tau_1, \tau_2 &\sim\textsf{Normal}(x_i,\tau_1^{-1}), & \text{if } \{i, i+1\}\notin C,\\
    x_{i+1}\mid x_i, \tau_1, \tau_2 &\sim\textsf{Normal}(x_i,\tau_2^{-1}), & \text{if } i\in C\text{ or } i + 1\in C.
\end{align*}
This is a simplified ARW1 with only two precisions: one precision for transitions not involving conflict years, $\tau_1$, and
one precision for transitions involving conflict years, $\tau_2$. As discussed in Section \ref{sec:rwanda_ex}, we expect that $\tau_1>\tau_2$. It follows that $\bx\mid \tau_{1},\tau_{2}\sim \textsf{Normal}(0,Q^-)$, where $Q$ is a special case of the general ARW1 precision matrix. We refer to this as a conflict adaptive random walk 1.

For this conflict ARW1 we can simplify $Q$ to $Q = \sum_{l=1}^2\tau_l R_l$, where for $l\in\{1,2\}$, $R_l=D_1-W_l$, where
\[
W_{l,{ij}}=
    \begin{cases}
      I(|i-j|=1\ \text{and}\ \{i, j\}\notin C), & \text{if}\ l=1, \\
      I(|i-j|=1\ \text{and}\ i\in C\text{ or } j\in C), & \text{if}\ l=2,
    \end{cases}
\]
and
$D_l=\text{diag}\left(\sum_{j=1}^{N}W_{l, {1j}},\dots, \sum_{j=1}^{N}W_{l, {Nj}}\right)$. Note that when $\tau_1 =\tau_2$ this random effect specification reduces to a RW1 as presented in Section \ref{sec:rw1}. In particular, in this case $Q = \tau_1( R_1 + R_2)$ where $R_1 + R_2$ is the structure matrix of a random walk 1.

\subsubsection{An Adaptive ICAR for Multiple Countries}
Suppose we are specifying a structured spatial random effect for $N$ regions at the Admin1 level with no islands, which are nested within $M$ countries. For neighboring Admin1 regions $i\sim j$, let $A_{ij}$ be an indicator that $i$ and $j$ are nested within the same country. Suppose for neighboring Admin1 regions $i\sim j$ we specify
\begin{align*}
    x_i-x_j\mid \tau_{1}, \tau_2 &\sim\textsf{Normal}(0,\tau_1^{-1}), & \text{ if } A_{ij}=1, \\
    x_i-x_j\mid \tau_{1}, \tau_2 &\sim\textsf{Normal}(0,\tau_2^{-1}), & \text{ if } A_{ij}=0.
\end{align*}
This is a simplified AICAR with only two precisions: one precision for neighboring Admin1 regions within the same country, $\tau_1$, and
one precision for neighboring Admin1 regions between different countries, $\tau_2$. As discussed in Section \ref{sec:multicountry_ex}, we expect that $\tau_1>\tau_2$. It follows that $\bx\mid \tau_{1},\tau_{2}\sim \textsf{Normal}(0,Q^-)$, where $Q$ is a special case of the general AICAR precision matrix. We refer to this as a multi-country AICAR.

For this multi-country AICAR we can simplify $Q$ to $Q = \sum_{l=1}^2\tau_l R_l$, where for $l\in\{1,2\}$, $R_l=D_1-W_l$, where
\[
W_{l,{ij}}=
    \begin{cases}
      I(i\sim j \text{ and } A_{ij}=1), & \text{if}\ l=1, \\
      I(i\sim j \text{ and } A_{ij}=0), & \text{if}\ l=2,
    \end{cases}
\]
and
$D_l=\text{diag}\left(\sum_{j=1}^{N}W_{l, {1j}},\dots, \sum_{j=1}^{N}W_{l, {Nj}}\right)$. Note that when $\tau_1 =\tau_2$ this random effect specification reduces to an ICAR as presented in Section \ref{sec:icar}. In particular, in this case $Q = \tau_1( R_1 + R_2)$ where $R_1 + R_2$ is the structure matrix of an ICAR.

\subsection{Connections to Previous Work}
Variants of the general first order AGMRFs introduced in Section \ref{sec:general_agmrfs} have been used before, typically for the purpose of constructing flexible and locally adaptive curve fitting methods in applications where shocks in the curves being fit are \textit{unknown} \citep{Lang_2002, Yue_2010, Rue_2005, Faulkner_2018, Faulkner_2019, Brewer_2007, Reich_2008}. We note that this is closely related to change point detection in the time series literature \citep{Aminikhanghahi_2017} and wombling in the spatial statistics literature \citep{Banerjee_2003, Lu_2005, Lu_2007, Carlin_2007, Heaton_2014}, where the goal is to \textit{identify} unknown shocks or regions of rapid change in curves. In the previous approaches, the models in \eqref{eq:arw1} and \eqref{eq:aicar} are used directly, with priors placed directly on all of the precision parameters, with the exception of \cite{Brewer_2007} and \cite{Reich_2008}, who restrict the AICAR in \eqref{eq:aicar} by reparameterizing the precision parameters as $\tau_{ij}=\tau_i\tau_j$, where $\tau_i,\tau_j$ are region-level precision parameters. While this parameterization reduces the number of precision parameters in the AICAR down to $N$, this is still a large number which requires the precision parameters to have smoothing priors of their own. 

\cite{Lang_2002, Yue_2010} induce dependence between these precision parameters by letting them follow GMRFs on the log scale. \cite{Rue_2005}, \cite{Faulkner_2018}, and \cite{Faulkner_2019} place priors on the precision parameters with the intention of marginalizing them out to produce non-normal differences. In particular, \cite{Faulkner_2018} and \cite{Faulkner_2019} focus on the case where differences marginally have horseshoe priors \citep{Carvalho_2010}. Due to the large number of precision parameters, these models are in most cases not amenable to fitting with INLA. Further, using MCMC to fit the models in \cite{Faulkner_2018} and \cite{Faulkner_2019} inherits the difficulties of using MCMC to sample from models with horseshoe priors \citep{Piironen_2017}, although this can be avoided by using priors with lighter tails. 

In contrast, in this work, we restrict the general first order AGMRFs as we have a priori knowledge of the location of the shocks in the curves we are fitting. This lets us work with a small number of precision parameters, allowing us to specify more substantive priors and fit the model in INLA, as we discuss in the following section. However, the prior information available in some applications may not always be enough to restrict an AGMRF to a small number of precision parameters. In such cases, MCMC or alternative, more flexible, deterministic Bayesian approximations, such as Template Model Builder \citep{Kristensen_2016}, would be necessary for model fitting.

Another approach to developing adaptive random effects models would be to consider adaptive generalizations of other spatial or temporal random effects. For areal spatial data, another common spatial random effect is the simultaneous autoregressive (SAR) model \cite[see e.g. Chapter 4 of][]{Banerjee_2003}. Compared to the adaptive generalizations of RWs and ICARs which have been proposed in the literature, there has been minimal work proposing adaptive generalizations of the SAR model \citep{Mukherjee_2014}.

In the Rwanda application, we have changed the smoothed direct model to account for events that have caused a shock in child mortality. This is related to the problem of incorporating information on feed-back interventions into forecasts in the dynamic linear models literature \citep{West_2006}. The dynamic linear models literature is motivated by providing more accurate forecasts in the future, whereas we are motivated by providing more accurate retrospective estimates of U5MR in the Rwanda application. 

In the multi-country application, we estimate U5MR at the subnational level, while leveraging the fact that subnational regions are nested within countries. However, it is sometimes desirable to simultaneously provide smoothed estimates of U5MR at the subnational and national level that are coherent to some extent. Methods from multiresolution modeling could be used for this purpose \citep{Ferreira_2007}; in particular multiscale random field models \citep{Ferreira_2005} would naturally fit into the smoothed direct modelling framework of \cite{Mercer_2015}.

\section{Scaling, Reparameterizations, Prior Choice, and Computation}
\label{sec:misc}
In this section, we describe various considerations for the use of our adaptive GMRFS in practice, including: scaling as in \cite{Sorbye_2014}, reparameterizations for interpretability, the choice of priors for hyperparameters, and computation.

\subsection{Scaling}
\label{sec:scaling}
Let $Q=\sum_{l=1}^2\tau_l R_l$ denote the precision matrix of one of the AGMRFs described in Section \ref{sec:usecases}. As these AGMRFs are improper, we need to worry about the interpretation of the precision parameters, which are dependent on the structure matrix $R_1+R_2$. In particular, we scale the precisions as in \cite{Sorbye_2014}. Let $\sigma^2(\bx)$ denote the geometric mean of the marginal variances of the elements of $\bx$ when setting $\tau_1=\tau_2=1$ (i.e. so that $Q=R_1+R_2$ is the structure matrix of a RW1 or ICAR). We work with the following scaled precision matrix $Q^* = \sum_{l=1}^2\tau_l R_l^*,$
where for $l\in\{1,2\}$, $R_l^*=R_l/\sigma^2(\bx)$. It then follows that when $\tau_1=\tau_2$, $1/\tau_1$ represents the approximate marginal variance of $\bx$, independent of the structure matrix $R_1+R_2$. This makes both interpretation and prior specification more straightforward.

\subsection{Reparameterizing for Interpretability}
\subsubsection{Reparameterization of AGMRFs}
Let $Q^* = \tau_1 R_1^* + \tau_{2} R_{2}^*$ denote the scaled precision matrix of one of the AGMRFs described in Section \ref{sec:usecases}. For both of the AGMRFs described in Section \ref{sec:usecases}, there is one precision, $\tau_1$, which we expect to be larger than the other, $\tau_2$: we expect the non-conflict precision to be larger than the conflict precision, and we expect the within-country precision to be larger than the between-country precision. We reparameterize $\tau_2$ such that $\tau_2=\tau_1\theta$, where $\theta\in(0,1]$, so that $Q^* = \tau_1 R_1^* + \tau_{1}\theta R_{2}^*=\tau_1(R_1^* + \theta R_{2}^*)$. 
We have that $\theta=\tau_2/\tau_1$, i.e., $\theta$ represents the ratio of the conflict precision to the non-conflict precision.

\subsubsection{A BYM2-Like Parameterization}
\label{sec:bym2like}
In the smoothed direct model introduced in Section \ref{sec:smoothed_direct}, when the structured area-level random effect was a RW1 or ICAR, we reparameterized the total area-level random effect, $\bb$, as a BYM2 following \cite{Riebler_2016}. Suppose instead our total area-level random effect is given by $\bb=\bv+\bx$, where $\bv$ is an unstructured random effect, and $\bx$ is one of the AGMRFs described in Section \ref{sec:usecases}, scaled as described in Section \ref{sec:scaling}. We now develop a BYM2-like parameterization of this total area-level random effect.
Let 
\[
\bb = \frac{1}{\sqrt{\tau_b}}\left(\sqrt{1-\phi}\bv +\sqrt{\phi}\bx^*\right)
\]
where $v_i\stackrel{iid}{\sim}\textsf{Normal}(0, 1)$,  $\bx^*\sim\textsf{Normal}\{0, (R_1^* + \theta R_{2}^*)^-\}$, and $\phi\in[0,1]$. Then $1/\tau_b$ represents the approximate marginal variance of the total area-level random effect $\bb$, and $\phi$ represents the proportion of this approximate variance attributed to the structured component, when $\theta=1$.

For computational purposes, we then reparameterize the $\bb$ to preserve sparsity as in \cite{Riebler_2016}. Let $\bw=(\bw_1^T\ \bw_2^T)^T$, where $\bw_1=\bb$ and $\bw_2 =\bx^*$. Then it follows that $\bw\sim\textsf{Normal}(0, S^-),$
where 
\[
S= 
\begin{pmatrix}
\frac{\tau_1}{1-\phi}I & -\frac{\sqrt{\phi\tau_1}}{1-\phi}I \\
-\frac{\sqrt{\phi\tau_1}}{1-\phi}I & R_1^* + \theta R_{2}^* + \frac{\phi}{1-\phi} I
\end{pmatrix}.
\]

\subsection{Prior Choice}
When using the AGMRFs described in Section \ref{sec:usecases} in a BYM2 like parameterization as outlined in Section \ref{sec:bym2like}, we must specify priors for the hyperparameters $\tau_b$, $\theta$, and $\phi$. We use the penalized complexity (PC) prior framework of \cite{Simpson_2017} to specify these priors. 

\subsubsection{PC Prior for $\tau_b$} We first consider the PC prior for $\tau_b$, where we are shrinking to $\tau_b=\infty$, i.e. no area-level random effect. Conditional on $\theta$ and $\phi$, the PC prior for $\tau_b$ is the PC prior for a normal precision as derived in \cite{Simpson_2017}
\[p(\tau_b\mid \theta)=\frac{\lambda}{2}\tau_b^{-3/2}\exp\left(-\lambda \tau_b^{-1/2}\right).
\]
In this paper, we specify the PC prior for  $\tau_b$ such that $P(1/\sqrt{\tau_b} > 1) = 0.01$.

\subsubsection{PC Prior for $\phi$}
We now consider the PC prior for $\phi$, where we are shrinking to $\phi=0$, i.e. no structured area-level random effect. For ease of implementation, we specify the PC prior for $\phi$ conditional on $\theta = 1$. Conditional on $\theta = 1$, the PC prior for $\phi$ is derived in Appendix C of \cite{Riebler_2016}, although it is not difficult to re-derive the PC prior for $\phi$ under a different value of $\theta$ if desired. In this paper, we specify the PC prior for $\phi$ such that $P(\phi < 0.5) = 2/3$.


\cite{Riebler_2016} derived the PC prior for $\phi$ numerically, by first deriving the KL divergence between the ``flexible" model and the ``base" model, $KL(\phi)$, then approximating the KL divergence as a function of $\phi$ using a spline, and then approximating the Jacobian, $\partial KL(\phi)/\partial \phi$, using the spline approximation. While this numerical approximation to the PC prior is easy to use in INLA, it is difficult to use in other software for Bayesian inference, such as Stan \citep{Carpenter_2017} and Template Model Builder \citep{Kristensen_2016}. Thus, in Appendix B of the Supplementary Materials we provide an analytical formula for the  PC prior for $\phi$, which may be of independent interest to practitioners using the BYM2 model outside of INLA.

\subsubsection{PC Prior for $\theta$} 
We now consider the PC prior for $\theta$, where we are shrinking to $\theta=1$, i.e. $\tau_1=\tau_2$ in the original parameterization of the AGMRFs described in Section \ref{sec:usecases}. In Appendix B of the Supplementary Materials, we derive an analytical formula for the PC prior for $\theta$ that shrinks to $\theta=1$. In this paper, we specify the PC prior for $\theta$ such that $P(\theta < 0.75) = 0.75$. 

\subsection{Computation}
As the AGMRFs described in Section \ref{sec:usecases} are Gaussian conditional on a low number of hyperparameters, they can be used as random effects in models fit using INLA. In particular, we implemented the AGMRFs through the \texttt{rgeneric} functionality in the \texttt{R-INLA} package. Implementing our AGMRFs using INLA allows them to be readily incorporated into the \texttt{R} package \texttt{SUMMER} \citep{Li_2020}, which contains implementations of the smoothed direct model of \cite{Mercer_2015} and has been used to produce estimates of U5MR supported by the UN IGME \citep{Li_2019}.

\section{Simulations}
\label{sec:simulations}
Before applying our AGMRFs to the motivating applications, we will perform a simulation study to assess whether the smoothed direct model using our AGMRFs can improve upon the performance of the smoothed direct model described in Section \ref{sec:smoothed_direct}. This simulation study is designed to mimic the structure of the Rwanda data which we analyze in Section \ref{sec:rwanda_app}.

We simulate data from the model 
\[
y_i\mid \eta_i \sim \textsf{Normal}(\eta_i, V),
\]
where $\eta_i = \mu_i + b_i$
and  $b_i\sim \textsf{Normal}(0, \tau_{i}^{-1})$.
There are $N=30$ time points, with time points 9-15 designated as conflict time points, imitating the structure of the Rwanda data. We vary $\mu_i$, $\tau_i$, and $V$ as follows:
\begin{itemize}
    \item $V$ takes values in $\{1/75, 1/150, 1/300\}$, which are in the range of the design-based variances associated with the direct estimates from the Rwanda data, as seen in Figure 1 of the Supplementary Materials.
    \item $\tau_i$ is either $20$ for all time points, or $20$ for non-conflict time points and $10$ for conflict time points.
    \item $\mu_i$ takes on one of three trends, as seen in Figure 2 of the Supplementary Materials. We refer to the trends as constant (no change during across time points), level change (increased mean during the conflict time points), and triangle (mean increasing and then decreasing during the conflict time points), based on their shapes.
\end{itemize}

We simulate 100 data sets from this model for each parameter setting and fit two models to the simulated data: the smoothed direct model with a RW1 for the structured temporal random effect, which we refer to as the smoothed direct model, and the smoothed direct model with our proposed conflict ARW1 for the structured temporal random effect, which we refer to as the proposed model. Both models are mispecified except in the case where $\mu_i$ is constant in time and the random effect precision is constant in time. We evaluate the two model fits with root mean squared error (RMSE) $\left(\sqrt{\frac{1}{N}\sum_{i=1}^N(\eta_i-\hat{\eta}_i)^2}\right)$, 
deviance information criterion (DIC) \citep{Spiegelhalter_2002}, and average proper logarithmic scoring rule (LS) \citep{Gneiting_2007}. For each metric, a lower value represents better model performance.

\begin{figure}[!ht]
\centering
\includegraphics[width=0.75\linewidth]{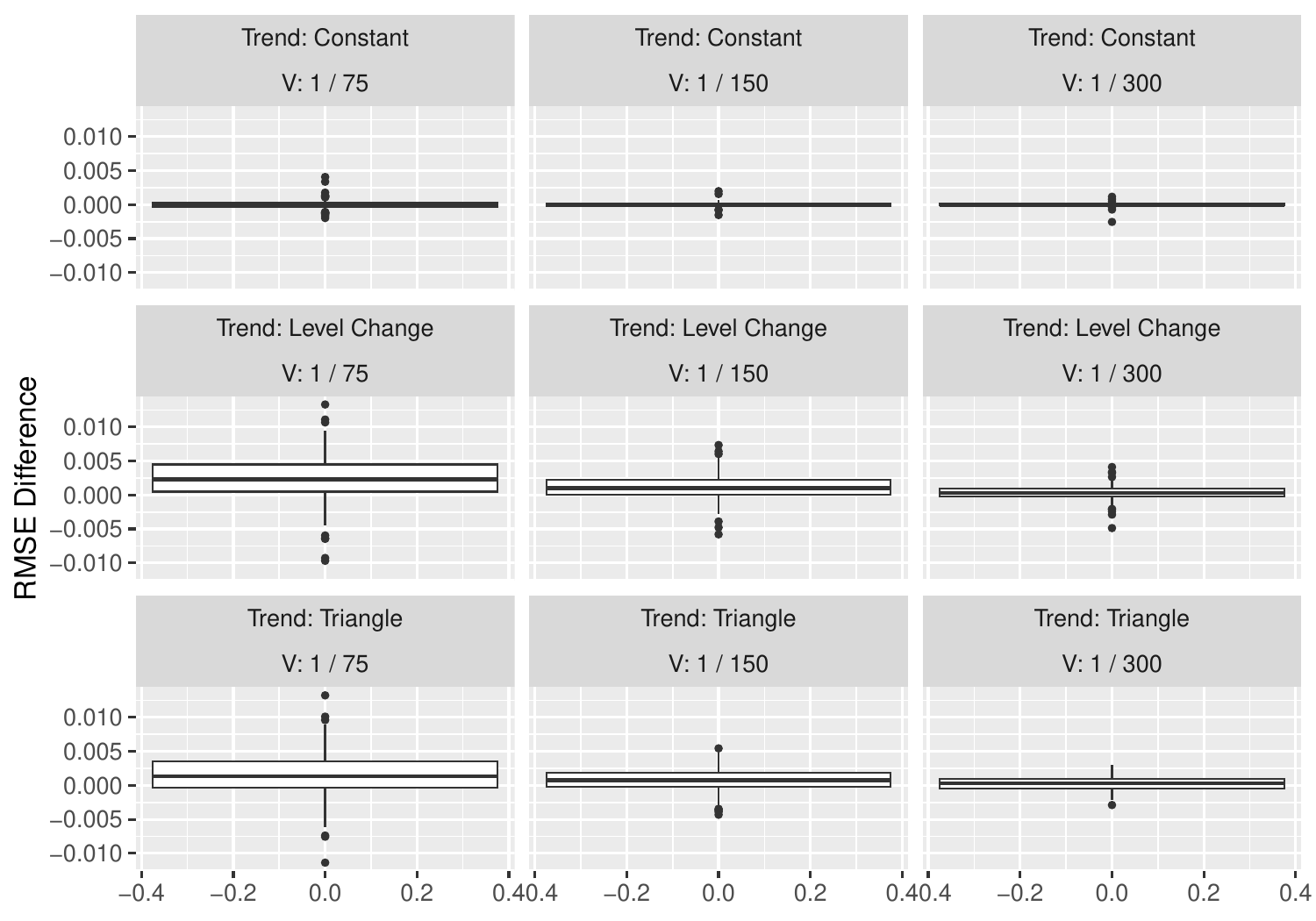}
\caption{Difference in RMSE between the smoothed direct model and the proposed model for simulation settings where $\tau_i$ is not the same for all time points. Positive difference indicates that the proposed model is outperforming the smoothed direct model.}
\label{fig:rmse_logit_risk_diff_plots_unequal}
\end{figure}

\begin{figure}[!ht]
\centering
\includegraphics[width=0.75\linewidth]{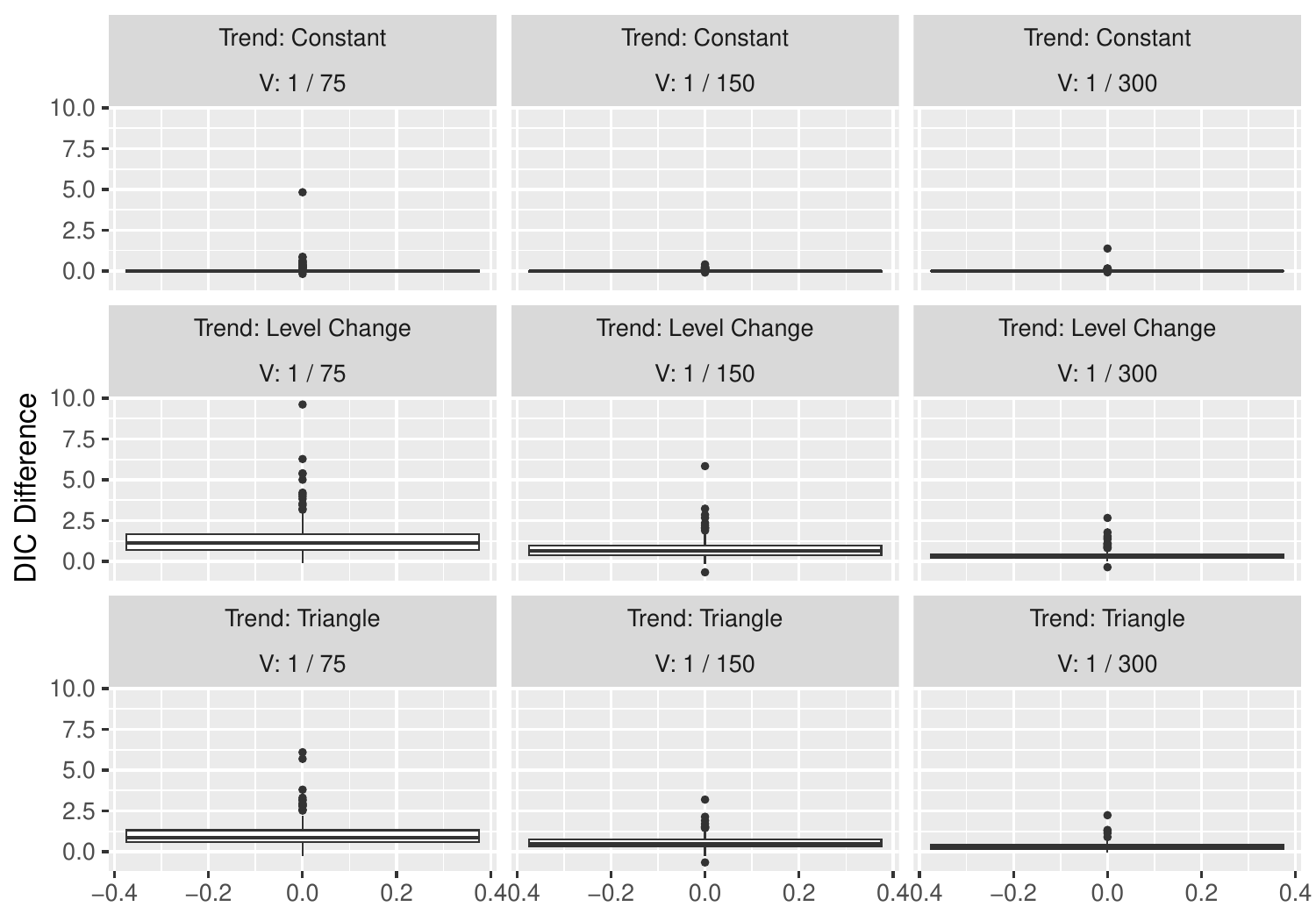}
\caption{Difference in DIC between the smoothed direct model and the proposed model for simulation settings where $\tau_i$ is not the same for all time points. Positive difference indicates that the proposed model is outperforming the smoothed direct model.}
\label{fig:dic_diff_plots_unequal}
\end{figure}

\begin{figure}[!ht]
\centering
\includegraphics[width=0.75\linewidth]{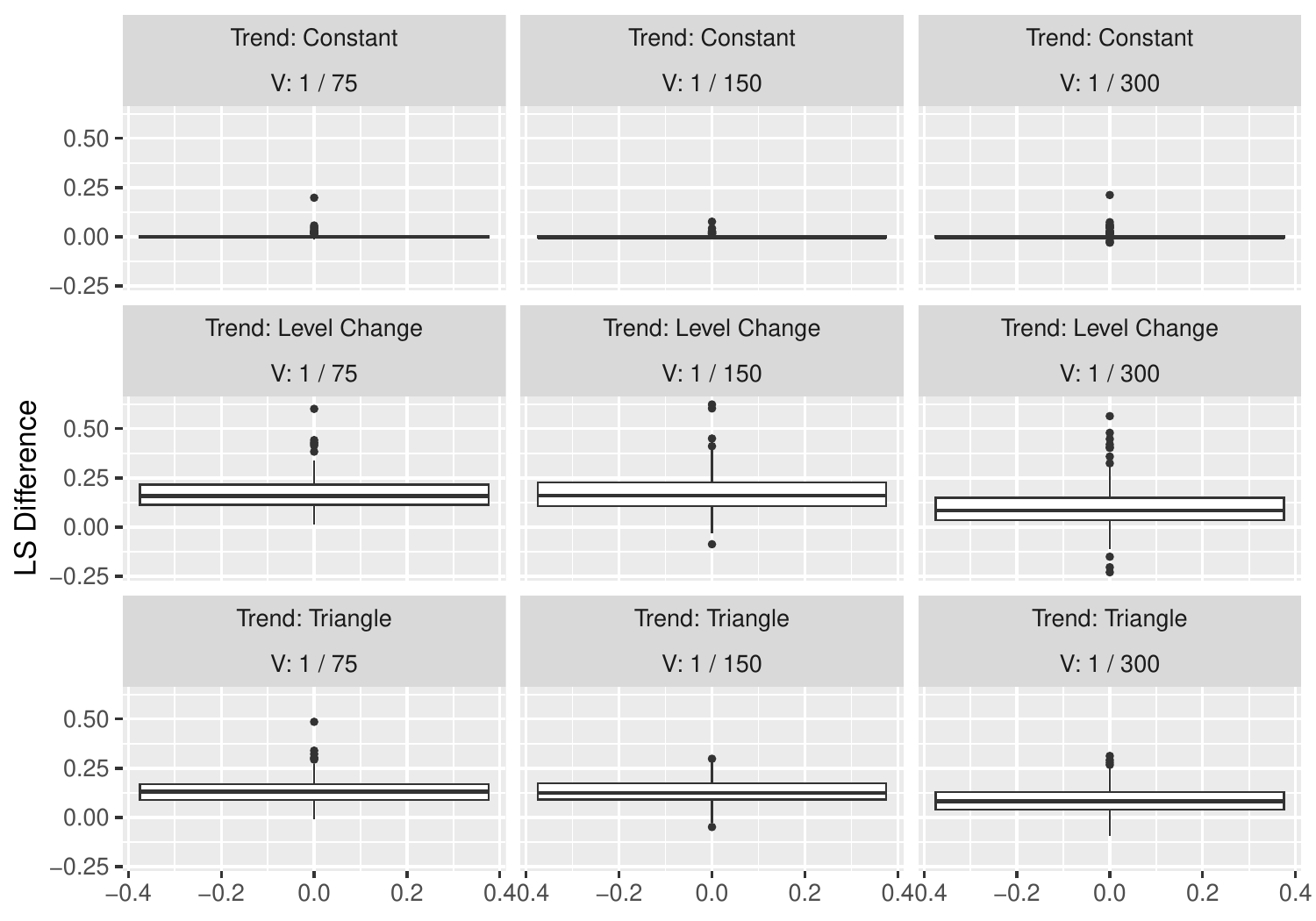}
\caption{Difference in LS between the smoothed direct model and the proposed model for simulation settings where $\tau_i$ is not the same for all time points. Positive difference indicates that the proposed model is outperforming the smoothed direct model.}
\label{fig:cpo_diff_plots_unequal}
\end{figure}

In Figures \ref{fig:rmse_logit_risk_diff_plots_unequal}, \ref{fig:dic_diff_plots_unequal}, and \ref{fig:cpo_diff_plots_unequal} we plot the difference in RMSE, DIC and LS, respectively, between the smoothed direct model and the proposed model, for the simulation settings where $\tau_i$ is not the same for all time points. 
As lower values of each metric represents better model performance, a positive difference in each metric indicates that the proposed model is outperforming the smoothed direct model. 
Across all metrics we find that the models have similar performance under the constant trend, and the proposed model outperforms the smoothed direct model under non-constant trends. Further, we see larger differences between the models as the $V$, which correspond to the design-based variances, decrease. Thus, we see that the potential for the proposed model to provide improvements over the smoothed direct model lies in the underlying curve having abrupt shocks, as in the case of the level change and triangle trends, as desired.

Plots for the the simulation settings where $\tau_i$ is the same for all time points show similar results, and are presented in Appendix C of the Supplementary Materials along with plots of the raw metrics across all simulation settings. 



\section{Estimation of U5MR at the National Level in Rwanda}
\label{sec:rwanda_app}
In this section, we estimate U5MR at the national level in Rwanda from 1985 through 2015, the last year with a DHS survey, and predict U5MR from 2016 through 2019. In Appendix E of the Supplementary Materials we describe the multi-country application. For this application, we slightly alter the smoothed direct model presented in Section \ref{sec:smoothed_direct}, adding a linear slope in time to the mean model $\eta_{is}$ to capture large scale temporal trends:
\begin{equation*}
    \eta_{is} =\mu + i\beta + v_i + x_i + \nu_s.
\end{equation*}
We use two models to estimate U5MR: this altered smoothed direct model with a RW1 for the structured temporal random effect, which we refer to as the smoothed direct model, and this altered smoothed direct model with our proposed conflict ARW1 for the structured temporal random effect, which we refer to as the proposed model. For the conflict ARW1, we categorize 1993-1999 as conflict years following UN IGME. We limit the direct estimates for each survey to those from the 15 years prior to when the survey was conducted. We note here that we explored fitting variants of both the smoothed direct model and the proposed model with separate intercepts for conflict and non-conflict periods, and the results did not substantively change.

In Table 1 of the Supplementary Materials, we summarize the posterior for the  various parameters in each model. The posterior summaries for $\mu$ are all larger under the proposed model. The posterior summaries for $\beta$ and $\phi$ are comparable between models. The posterior summaries for $\nu_s$, the survey random effects, are also comparable between models, and are all centered around $0$ except for the effect for the 2008 survey, which has a $95\%$ credible interval of $(-0.16, 0.00)$. Examining figure \ref{fig:rwanda_direct}, we see that the direct estimates in the 90s from the 2008 survey are a bit lower than those from the 2000 and 2005 surveys, which explains the posterior for $\nu_{2008}$ being negative across the two models.

The posterior summaries for $\tau$ are all larger under the proposed model. As $\phi/\tau$ represents the approximate marginal variance of the total temporal random effect attributed to the structured component when $\theta=1$, a larger $\tau$ means that the marginal variance of the structured component for non-conflict time periods in the proposed model is smaller than the marginal variance of the structured component in the smoothed direct model.

\begin{figure}[!ht]
\centering
\includegraphics[width=0.75\linewidth]{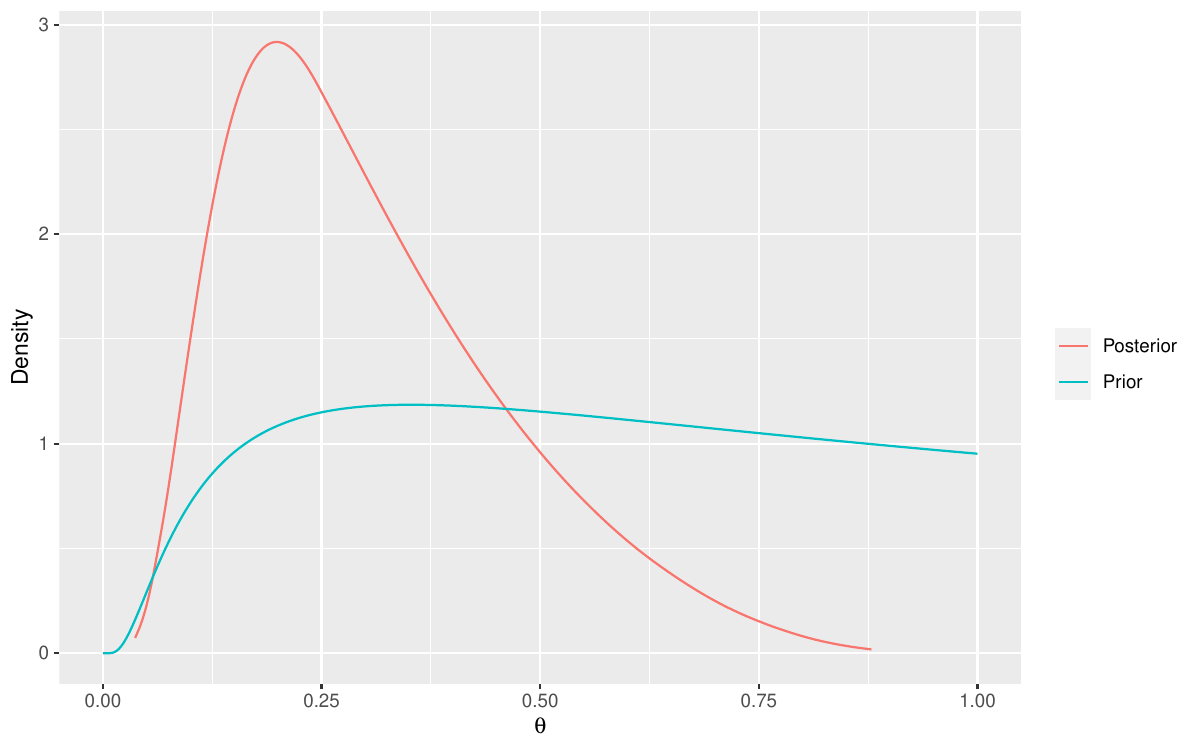}
\caption{Comparison of prior and posterior density for $\theta$ in the Rwanda application.}
\label{fig:theta_comp_rwanda}
\end{figure}

Focusing on $\theta$, in Figure \ref{fig:theta_comp_rwanda} we plot the prior and posterior for $\theta$ from the proposed model. We see that the prior is relatively flat from 0.2 to 1, with $95\%$ of the prior mass lying in $[0.09, 0.97]$. The posterior median is $0.28$, and the $95\%$ credible interval for $\theta$ is $[0.09, 0.68]$. The posterior places very little mass close to 1, and is relatively concentrated around the mode of $0.20$. Thus under the proposed model, we estimate that the conflict precision is roughly $20$-$30\%$ of the non-conflict precision in the conflict ARW1.

\begin{figure}[!ht]
\centering
\includegraphics[width=0.99\linewidth]{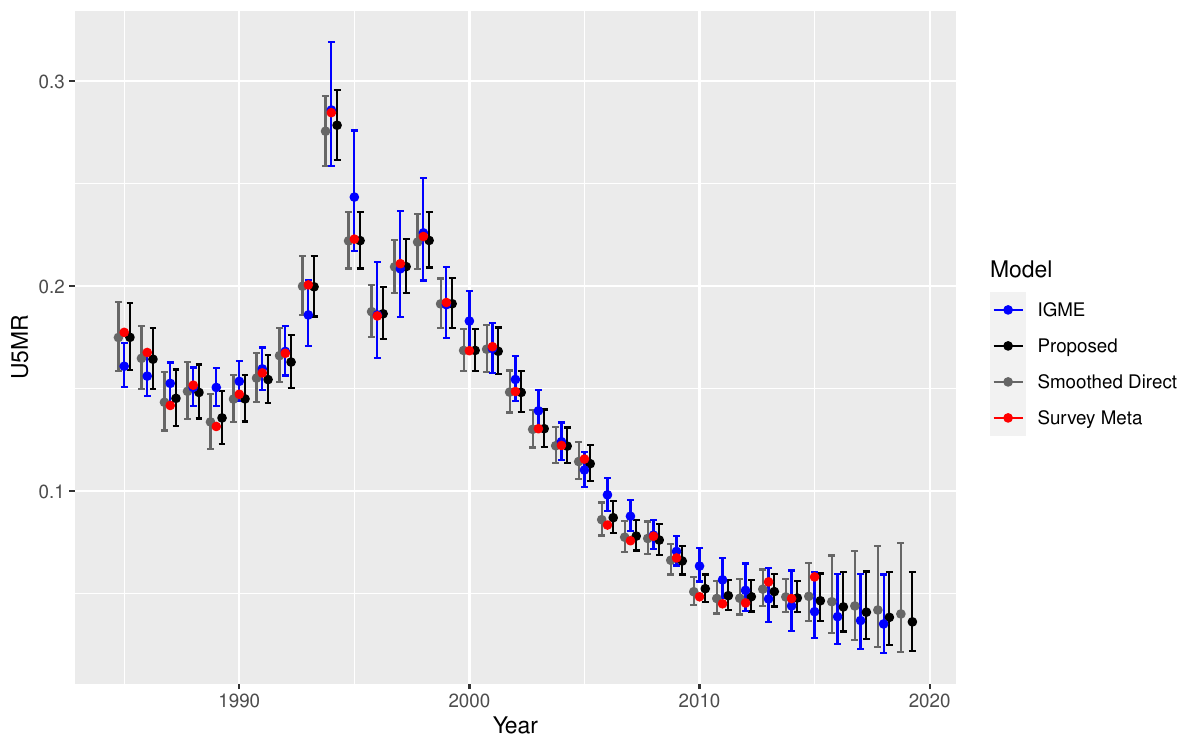}
\caption{Comparison of U5MR estimates from the smoothed direct model, the proposed model, UN IGME, and meta-analysis estimator of U5MR based on the direct estimates. 90\% credible intervals are shown for the smoothed direct and proposed model, 90\% uncertainty intervals are shown for UN IGME.}
\label{fig:rwanda_estimates_1}
\end{figure}

In Figure \ref{fig:rwanda_estimates_1}, we plot U5MR estimates from the smoothed direct model and the proposed model, in addition to the UN IGME and meta-analysis estimates discussed in Section \ref{sec:rwanda_ex} for reference. 90\% credible intervals are shown for the smoothed direct and proposed model and 90\% uncertainty intervals are shown for UN IGME. The estimates from the smoothed direct and proposed models differ the most at two time periods: 1994 and 2016-2019. 1994 was the peak of the Rwandan civil war, when the Rwandan genocide occurred, which lead to a shock child mortality. The proposed model is able to avoid oversmoothing this time period, compared to the smoothed direct model, leading a larger estimate of U5MR that is closer to the meta analysis estimate.

The last year in which a survey was conducted was 2015, so we predict U5MR from 2016-2019. We see that the proposed model has much narrower prediction intervals over this time period than the smoothed direct model. The smoothed direct model uses a RW1 for the structured temporal random effect, which has a single variance for all temporal transitions. Thus, if transitions not involving conflicts truly have a smaller variance than transitions involving conflicts, the single variance parameter in the smoothed direct model has to deal with with this behavior by assuming a value somewhere in between the variance of the two types of transitions. This means that transitions not involving conflicts, which includes 2016-2019, will have a larger variance than under the smoothed direct model than under the proposed model, leading to narrower prediction intervals under the proposed model.

\section{Conclusions}
In this article, we developed a class of adaptive GMRFs which can incorporate knowledge of expected shocks in mortality, while still allowing information to be borrowed across space and time. Methodologically, an important next step will be to develop spatio-temporal models that incorporate our adaptive GMRFs. A natural first step would be to utilize the spatio-temporal interaction framework of \cite{Knorr_2000}. For example, if one wanted to estimate U5MR across multiple countries over multiple time periods, one could use a type IV interaction from \cite{Knorr_2000}, where the spatial component is a multi-country AICAR and the temporal component is a non-adaptive GMRF, like a RW1.

In the Rwanda application, we focused in this article on developing a national model for U5MR from 1985-2015 that did not oversmooth during conflict years. However, we would ultimately like to develop a \textit{subnational} model for U5MR from 1985-2015. This poses challenges, as earlier surveys that provide the most information for conflict years do not have GPS information available for sampled households, unlike later surveys. The only geographic information these earlier surveys have for sampled households is administrative region at the time of the survey, and the administrative division of Rwanda changed in 2006. Thus, future work will need to develop a spatial model that harmonizes these different administrative divisions. This new spatial model could then be combined with the conflict ARW1 to develop a subnational model for U5MR from 1985-2015.

\section*{Software}
Software reproducing the simulation and applications is available at the following github link: \url{https://github.com/aleshing/AGMRF_code}.

\section*{Acknowledgments}
We thank Abel Rodriguez for providing references to related works, Andrea Riebler and Geir-Arne Fuglstad for discussions regarding the derivations of PC priors, and Rebecca Steorts, Ryan Janicki, Eric Slud, and Tommy Wright for helpful comments.

\bibliographystyle{biorefs}
\bibliography{refs}

\end{document}


\maketitle

\appendix

\tableofcontents

\section{Higher Order Adaptive Gaussian Markov Random Fields}
Extensions to higher order AGMRFs are straightforward mathematically, but not so much conceptually, as we will now briefly illustrate with an adaptive random walk of second order. Suppose we have $N$ time periods and we are specifying a structured temporal random effect $\bx=(x_1,\dots,x_N)$. Suppose for $i=1,\dots, N-2$ we specify 
\begin{equation*}
x_{i+2}-x_{i+1}\mid x_{i+1}, x_{i}, \tau_i \sim\textsf{Normal}(x_{i+1}-x_{i},\tau_i^{-1})
\end{equation*}
where $\tau_i$ is the precision for the $i$th second order difference. This leads to an adaptive version of the second-order random walk \citep[see e.g. page 110 of][]{Rue_2005}. 

As with the ARW1, as is this model has a large number of precision parameters, which makes it not parsimonious and computationally difficult to fit. However, because the model is defined using second order differences, rather than first order differences, it is not clear how one should go about specializing the model as in Section 3.2 of the main paper. In particular, suppose we were trying to specialize this adaptive second-order random walk to the Rwanda application as in Section 3.2.1 of the main text. For the first-order random walk, we were able to use the interpretation of first order differences to reduce the parameter space down to two precisions for differences involving conflicts and not involving conflicts. It is not clear to the authors how to use the interpretation of second order differences to reduce the parameter space. This difficulty compounds when moving to even higher order GMRFs.

\section{Prior Appendix}
In this appendix, we derive analytical formulas for the PC priors for the BYM2 $\phi$ parameter and the $\theta$ parameter in our proposed AGMRFs.

\subsection{Analytical Formula for BYM2 $\phi$ PC Prior}
In this appendix, we derive an analytical formula for the PC prior for $\phi$ that shrinks to $\phi=0$. This involves: 1) deriving the KL divergence between our ``flexible" model, where $\phi\in[0,1]$, and our ``base" model, where $\phi=0$, 2) placing an exponential prior on a transformation of the KL divergence, and 3) transforming the prior on the KL divergence to a prior on $\phi$.

Let $Q_*$ be the scaled structure matrix for the structured component of the random effect. Following \cite{Riebler_2016}, we have that the KL divergence between the flexible and base models is 
\begin{align*}
d(\phi) &=\sqrt{2KL[\textsf{Normal}\{0, (1-\phi)I+\phi Q^-_*\} \ || \ \textsf{Normal}\{0, I \}] }\\
&=\sqrt{ \text{trace}\{(1-\phi)I+\phi Q^-_*\}-n-\log\{|(1-\phi)I+\phi Q^-_*|\}}.
\end{align*}
Let $\gamma_i$ denote the eigenvalues of $Q_*$, let $\tilde{\gamma}_i=1/\gamma_i$ if $\gamma>0$ and $\tilde{\gamma}_i=0$ otherwise. We can then simplify the distance $d(\phi)$ as follows:
\begin{align*}
    d(\phi)&=\sqrt{\text{trace}\{(1-\phi)I+\phi Q^-_*\}-n-\log\{|(1-\phi)I+\phi Q^-_*|\}}\\
    &=\sqrt{\left\{\sum_{i=1}^n1 + \phi(\tilde{\gamma}_i-1)\right\}-n-\log\left\{\prod_{i=1}^n1 + \phi(\tilde{\gamma}_i-1)\right\}}\\
    &=\sqrt{\phi\sum_{i=1}^n (\tilde{\gamma}_i-1)-\sum_{i=1}^n\log\left\{1 + \phi(\tilde{\gamma}_i-1)\right\}}.
\end{align*}

Using this expression for $d(\phi)$, the PC prior for $\phi$ is then 
\[
p(\phi)=\lambda\exp\{-\lambda d(\phi)\}|\partial d(\phi)/\partial\phi|,
\]
where 
\[
\partial d(\phi)/\partial\phi=\frac{\phi}{2d(\phi)}\sum_{i=1}^n\frac{(\tilde{\gamma}_i-1)^2}{1 + \phi(\tilde{\gamma}_i-1)}.
\]

\subsection{Analytical Formula for $\theta$ PC Prior}
In this appendix, we derive an analytical formula for the PC prior for $\theta$ that shrinks to $\theta=1$. This involves: 1) deriving the KL divergence between our ``flexible" model, where $\theta\in(0,1]$, and our ``base" model, where $\theta=1$, 2) placing an exponential prior on a transformation of the KL divergence, and 3) transforming the prior on the KL divergence to a prior on $\theta$.

We first need to derive the KL divergence between our flexible model and the base model. 
As both the base and flexible models are intrinsic, we will instead calculate the KL divergence in the $n-1$ subspace where we have taken away the singular portion of the Gaussians. In particular, let $\hat{R}_1^*$ and $\hat{R}_2^*$ denote $R_1^*$ and $R_2^*$ after removing the first row and column, although we could remove in general the $r$th row and column and the result would stay the same. Then we can now derive the KL divergence between our flexible model and base model:
\begin{align*}
    d(\theta)&=
    \sqrt{
    2KL(
    \textsf{Normal}[0, \{\tau_1(\hat{R}_1^* + \theta \hat{R}_{2}^*)\}^{-1}]
    \ || \
    \textsf{Normal}[0, \{\tau_1(\hat{R}_1^* + \hat{R}_{2}^*)\}^{-1} ]
    ) 
    }\\ 
    &= \sqrt{\text{trace}\left\{(\hat{R}_1^* + \hat{R}_{2}^*)(\hat{R}_1^* + \theta \hat{R}_{2}^*)^{-1}\right\} - (n-1) - \log\left(\frac{|\hat{R}_1^* +  \hat{R}_{2}^*|}{|\hat{R}_1^* + \theta\hat{R}_{2}^*|}\right)}.
\end{align*}

Note that we can write $\hat{R}_1^* + \theta\hat{R}_{2}^*=\hat{R}_1^* + \hat{R}_{2}^* + \hat{R}_{2}^*(\theta - 1)=(\hat{R}_1^* + \hat{R}_{2}^*)\{I + (\hat{R}_1^* + \hat{R}_{2}^*)^{-1}\hat{R}_{2}^*(\theta - 1)\}$. Let $\varepsilon_1,\dots, \varepsilon_{n-1}$ denote the $n-1$ eigenvalues of $(\hat{R}_1^* + \hat{R}_{2}^*)^{-1}\hat{R}_{2}^*$. It follows that we can simplify the calculation of $d(\theta)$ as follows:
\begin{align*}
    d(\theta)&= \sqrt{\text{trace}\{(\hat{R}_1^* + \hat{R}_{2}^*)(\hat{R}_1^* + \theta \hat{R}_{2}^*)^{-1}\} - (n-1) - \log\left(\frac{|\hat{R}_1^* +  \hat{R}_{2}^*|}{|\hat{R}_1^* + \theta\hat{R}_{2}^*|}\right)}\\
    &= \sqrt{\text{trace}[\{I + (\hat{R}_1^* + \hat{R}_{2}^*)^{-1}\hat{R}_{2}^*(\theta - 1)\}^{-1}] - (n-1) + \log\left\{|I + (\hat{R}_1^* + \hat{R}_{2}^*)^{-1}\hat{R}_{2}^*(\theta - 1)|\right\}}\\
    &= \sqrt{\sum_{i=1}^{n-1}\frac{1}{1+(\theta - 1)\varepsilon_i} - (n-1) + \log\left\{\prod_{i=1}^{n-1}1+(\theta - 1)\varepsilon_i\right\}}\\
    &= \sqrt{\sum_{i=1}^{n-1}\frac{1}{1+(\theta - 1)\varepsilon_i} - (n-1) + \sum_{i=1}^{n-1}\log\left\{1+(\theta - 1)\varepsilon_i\right\}}.
\end{align*}

Placing an exponential prior on $d(\theta)$ and transforming to a prior on $\theta$, we find that the PC prior for $\theta$ is given by
\[p(\theta)= \lambda\exp\{-\lambda d(\theta)\}\left|\frac{\partial d(\theta)}{\partial \theta}\right|,
\]
where we can calculate the Jacobian as 
\begin{align*}
    \frac{\partial d(\theta)}{\partial \theta}&=\frac{\partial }{\partial \theta}\sqrt{d^2(\theta)}\\
    &=\frac{1}{2}\left\{d(\theta)\right\}^{-1}
    \frac{\partial d^2(\theta)}{\partial \theta}\\
    &=\frac{1}{2}\left\{d(\theta)\right\}^{-1}\sum_{i=1}^{n-1}\frac{(\theta - 1)\varepsilon_i^2}{\{1+(\theta - 1)\varepsilon_i\}^2}.
\end{align*}
Thus the PC prior for $\theta$ can be written as
\[p(\theta)= \frac{\lambda (1-\theta)}{2d(\theta)}\left[\sum_{i=1}^{n-1}\frac{\varepsilon_i^2}{\{1+(\theta - 1)\varepsilon_i\}^2}\right]\exp\{-\lambda d(\theta)\}.
\]

The user can specify the parameter $\lambda$ using a prior probability statement of the form $P(\theta<U)=\alpha$. As $P(\theta<U)=P\{d(\theta)>d(U)\}=\exp\{-\lambda d(U)\}=\alpha$, since $d(\theta)$ is a decreasing function of $\theta$ on $(0,1]$, this corresponds to using $\lambda = -\log(\alpha)/d(U)$.

\section{Simulation Appendix}
\begin{figure}[!ht]
\centering
\includegraphics[width=0.85\linewidth]{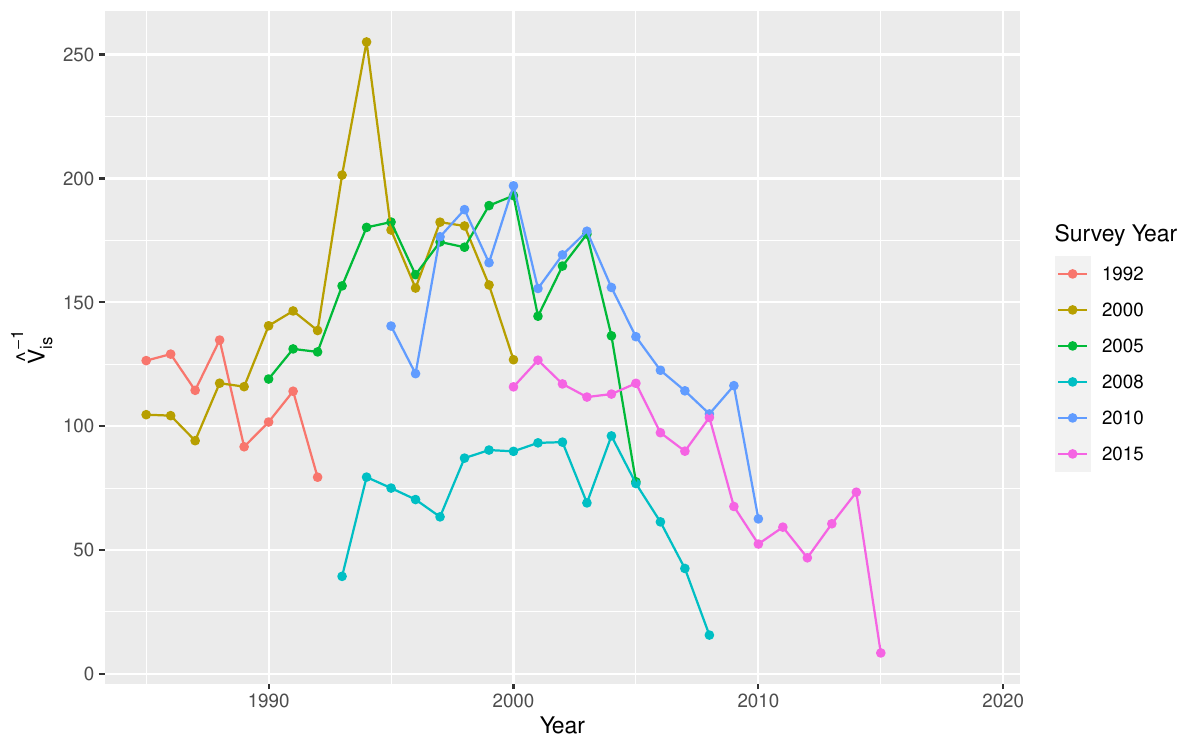}
\caption{Precisions associated with the direct estimates of logit-U5MR from the Rwanda data.}
\label{fig:survey_prec}
\end{figure}

\begin{figure}[!ht]
\centering
\includegraphics[width=1\linewidth]{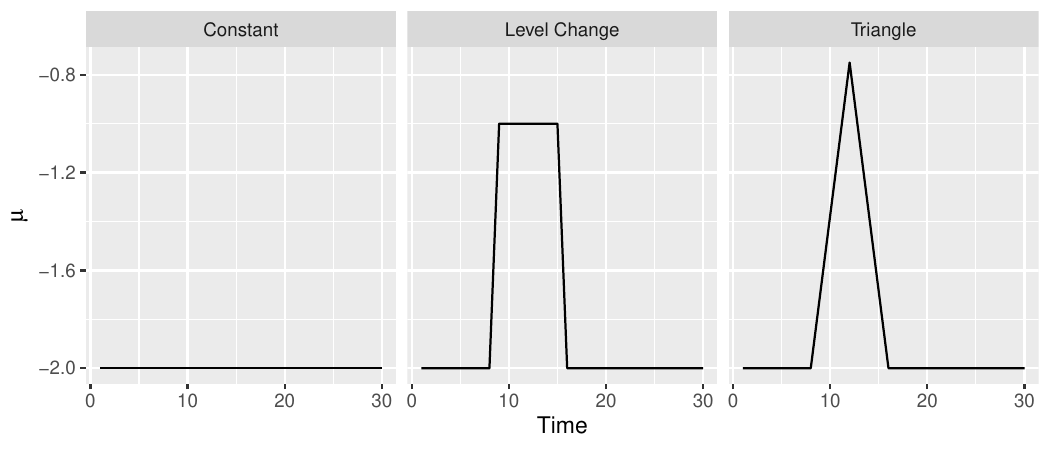}
\caption{The three trends used for $\mu_i$ to simulate data.}
\label{fig:trends}
\end{figure}

In this appendix we present additional plots from the simulation study carried out in Section 5 of the main text. In Figure \ref{fig:survey_prec} we plot $\hat{V}_{is}^{-1}$, the precisions associated with the direct estimates of logit-U5MR from the Rwanda data.
In Figure \ref{fig:trends} we plot the three different trends $\mu_i$ takes on in the simulation. 

\begin{figure}[!ht]
\centering
\includegraphics[width=0.65\linewidth]{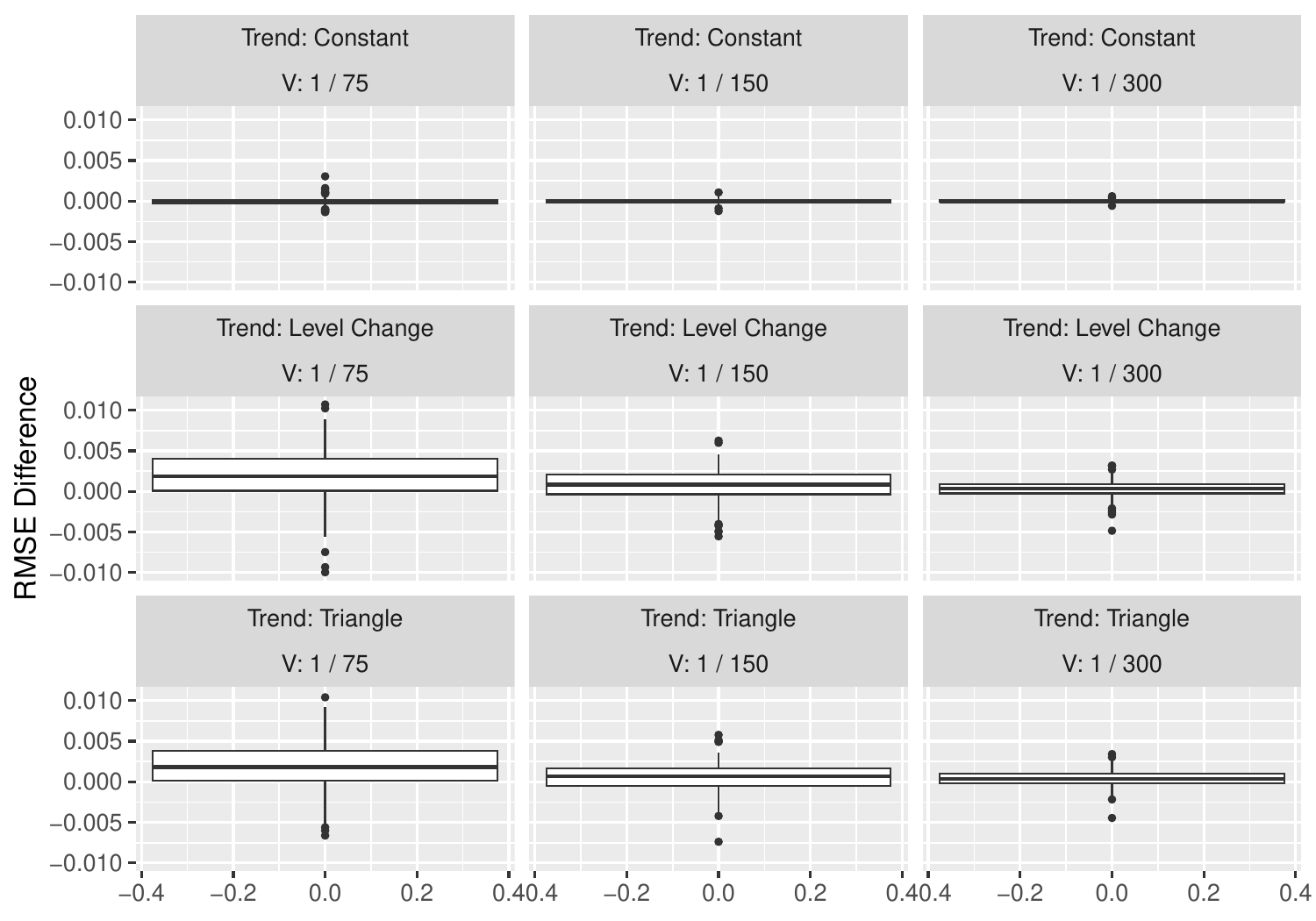}
\caption{Difference in RMSE between the smoothed direct model and the proposed model for simulation settings where $\tau_i$ is the same for all time points.}
\label{fig:rmse_logit_risk_diff_plots_equal}
\end{figure}

\begin{figure}[!ht]
\centering
\includegraphics[width=0.65\linewidth]{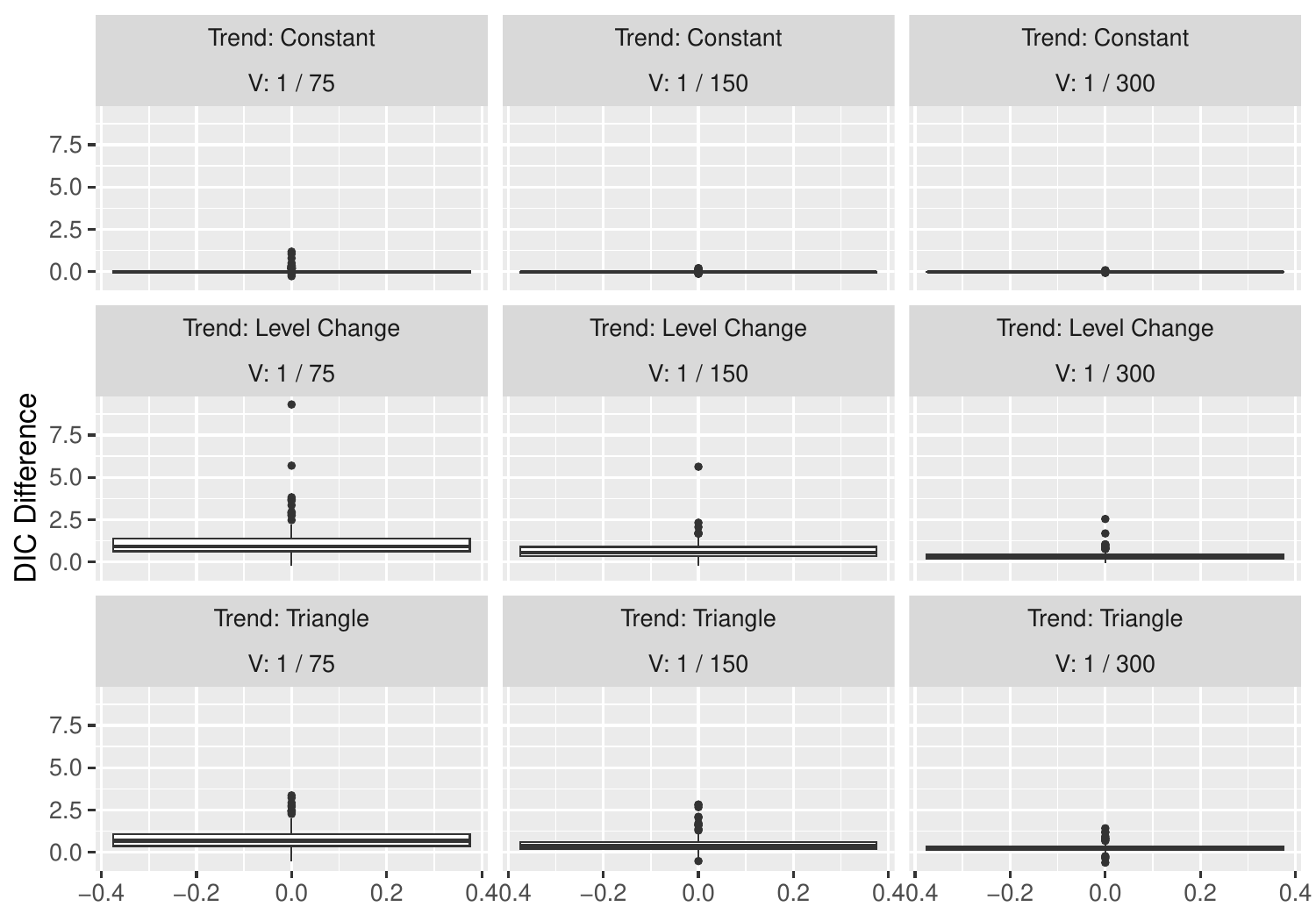}
\caption{Difference in DIC between the smoothed direct model and the proposed model for simulation settings where $\tau_i$ is the same for all time points.}
\label{fig:dic_diff_plots_equal}
\end{figure}

\begin{figure}[!ht]
\centering
\includegraphics[width=0.65\linewidth]{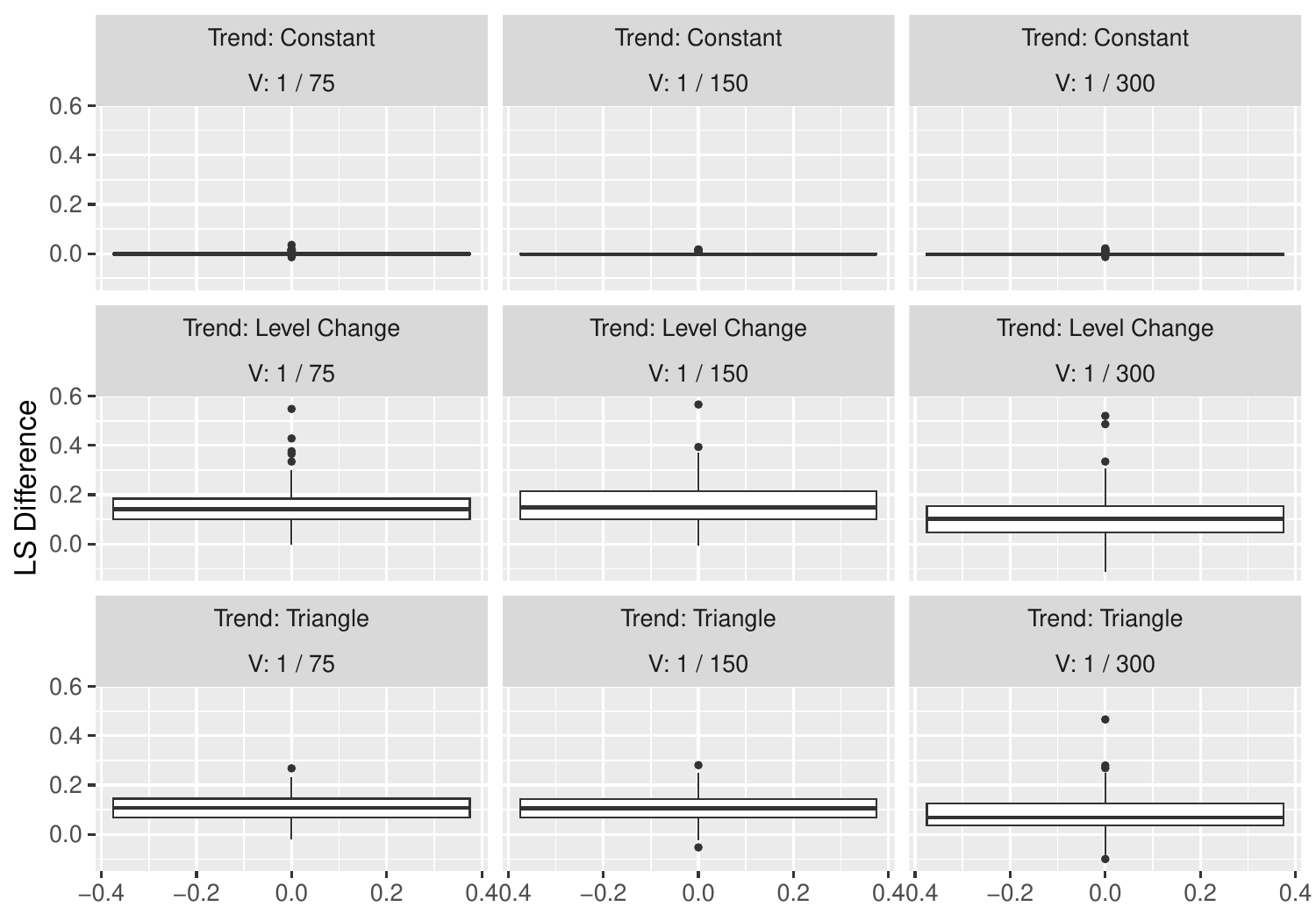}
\caption{Difference in LS between the smoothed direct model and the proposed model for simulation settings where $\tau_i$ is the same for all time points.}
\label{fig:cpo_diff_plots_equal}
\end{figure}

In Figures \ref{fig:rmse_logit_risk_diff_plots_equal}, \ref{fig:dic_diff_plots_equal}, and \ref{fig:cpo_diff_plots_equal} we plot the difference in RMSE, DIC and LS, respectively, between the smoothed direct model and the proposed model, for the simulation settings where $\tau_i$ is the same for all time points. The results in these settings closely mimic the results shown in Section 5 of the main text.

\begin{figure}[!ht]
\centering
\includegraphics[width=0.65\linewidth]{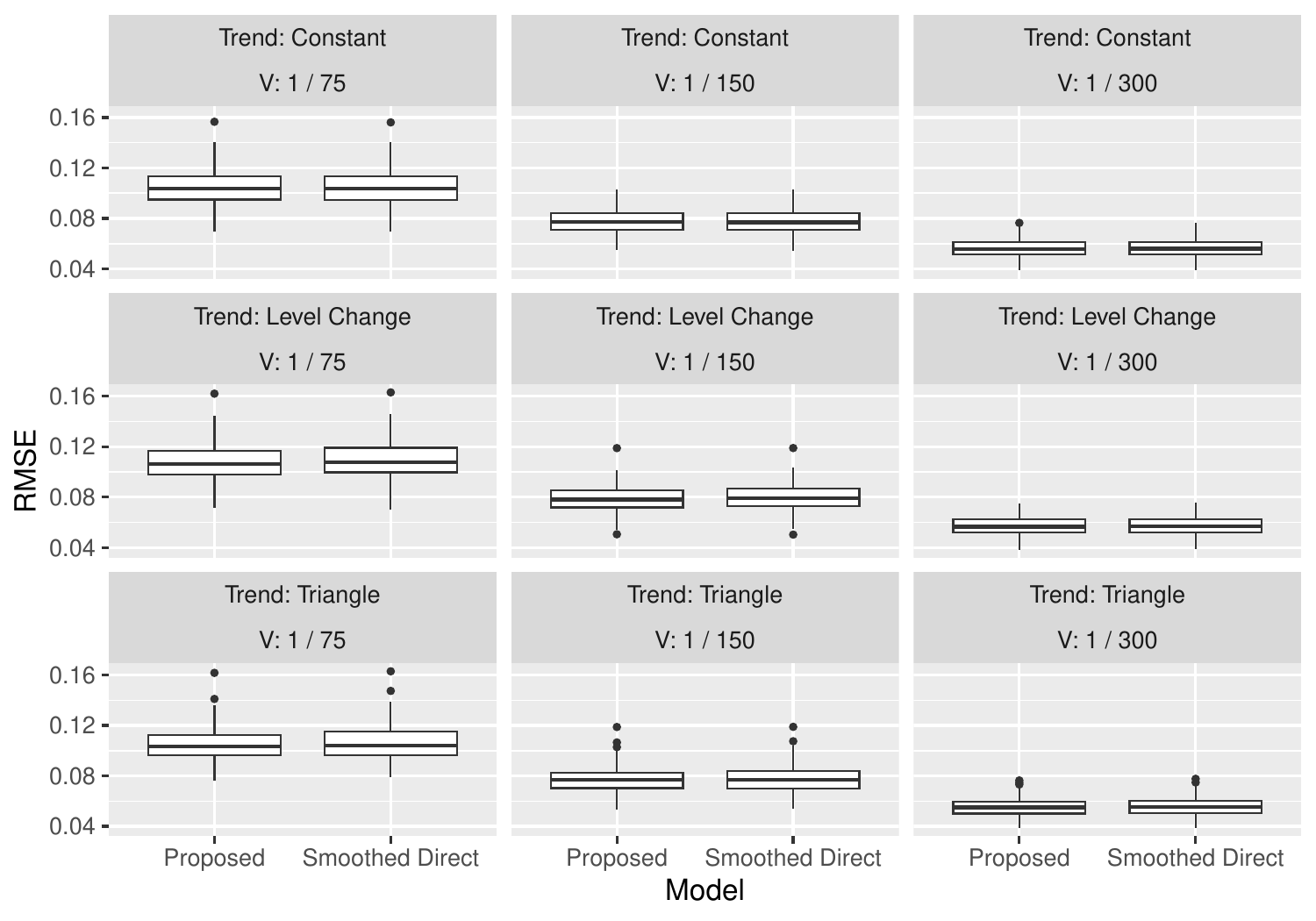}
\caption{RMSE for the proposed and smoothed direct models for simulation settings where $\tau_i$ is the same for all time points.}
\label{fig:rmse_logit_risk_plots_equal}
\end{figure}

\begin{figure}[!ht]
\centering
\includegraphics[width=0.65\linewidth]{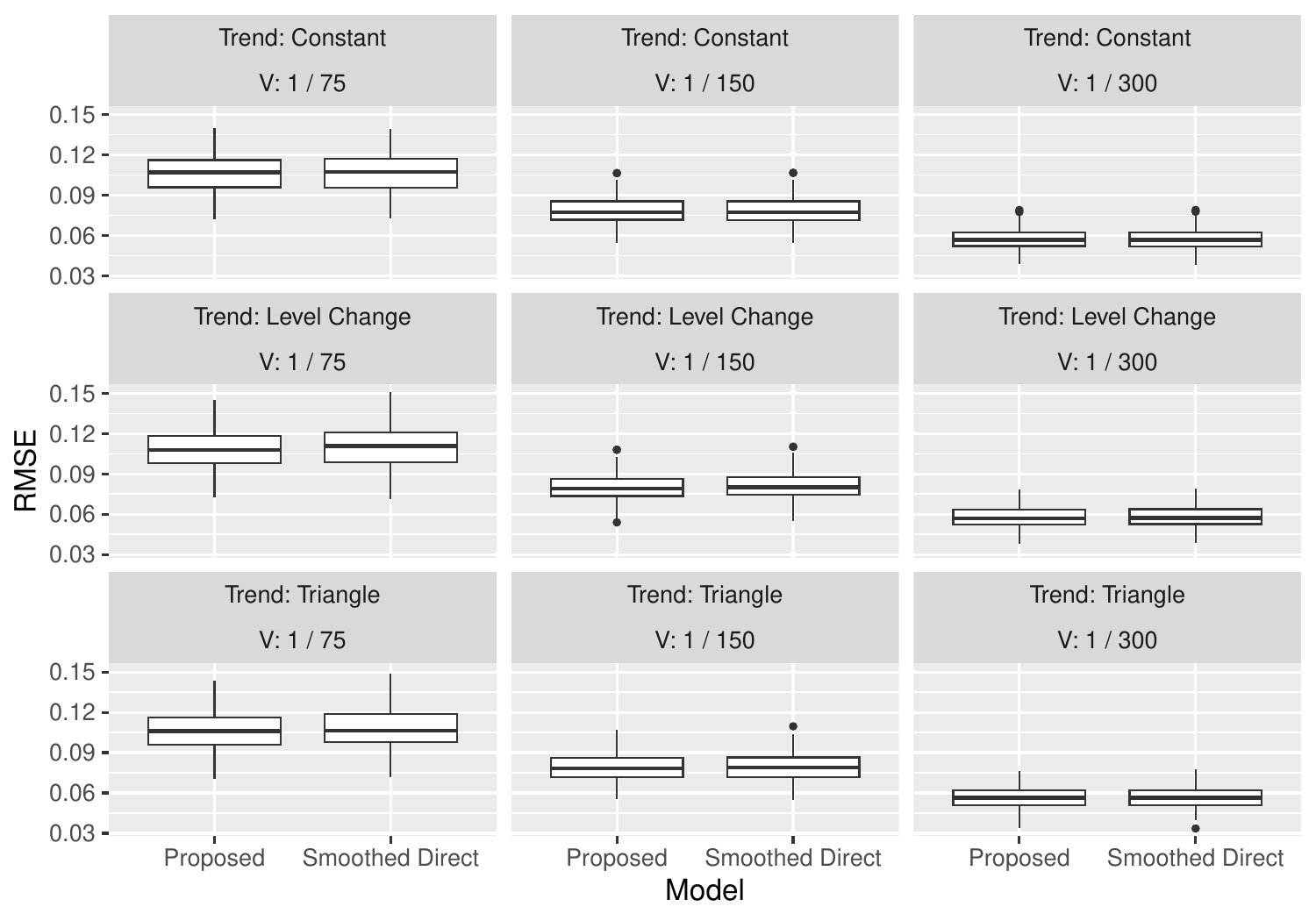}
\caption{RMSE for the proposed and smoothed direct models for  simulation settings where $\tau_i$ is not the same for all time points.}
\label{fig:rmse_logit_risk_plots_unequal}
\end{figure}

\begin{figure}[!ht]
\centering
\includegraphics[width=0.65\linewidth]{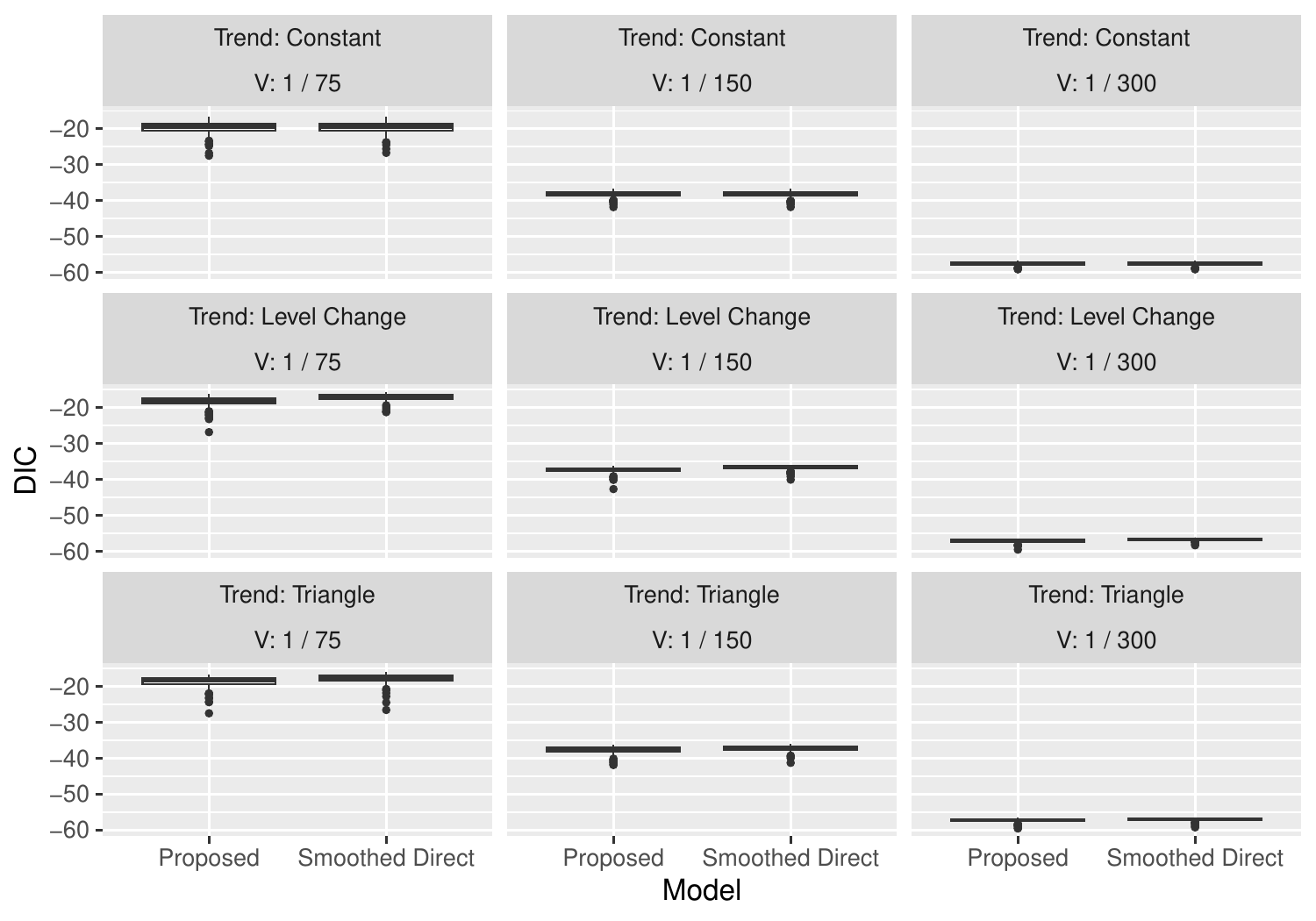}
\caption{DIC for the proposed and smoothed direct models for  simulation settings where $\tau_i$ is the same for all time points.}
\label{fig:dic_plots_equal}
\end{figure}

\begin{figure}[!ht]
\centering
\includegraphics[width=0.65\linewidth]{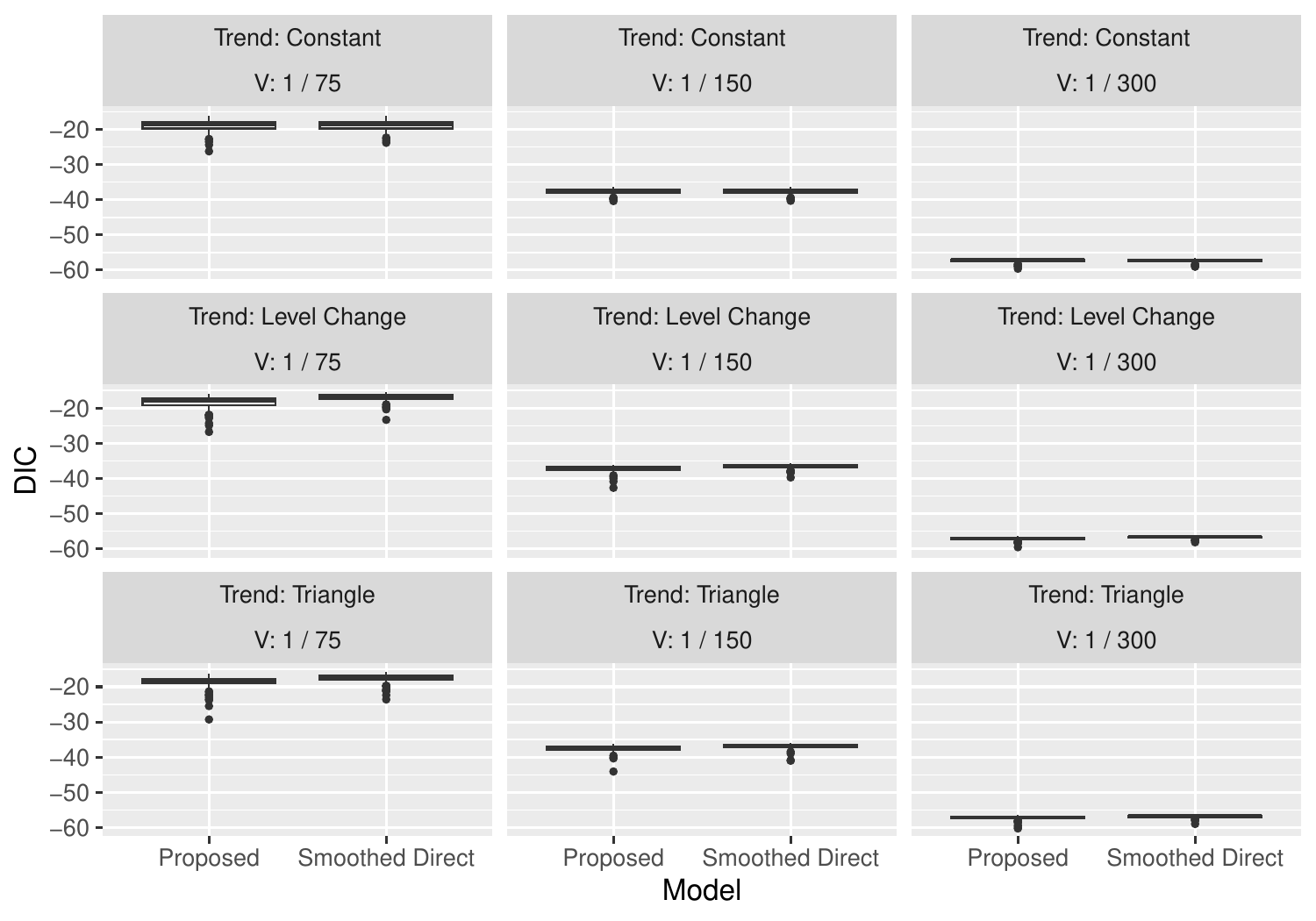}
\caption{DIC for the proposed and smoothed direct models for  simulation settings where $\tau_i$ is not the same for all time points.}
\label{fig:dic_plots_unequal}
\end{figure}

\begin{figure}[!ht]
\centering
\includegraphics[width=0.65\linewidth]{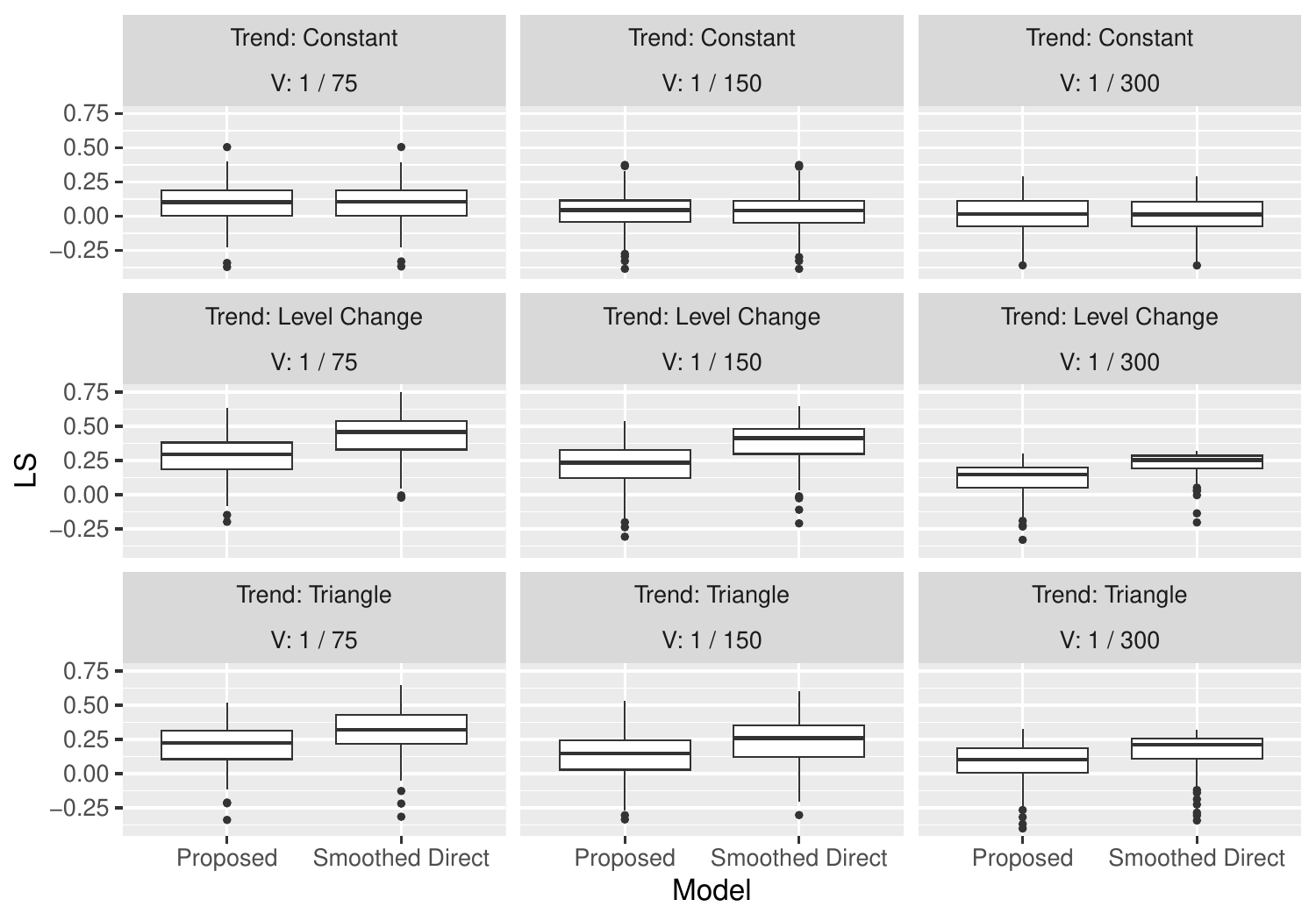}
\caption{LS for the proposed and smoothed direct models for  simulation settings where $\tau_i$ is the same for all time points.}
\label{fig:cpo_plots_equal}
\end{figure}

\begin{figure}[!ht]
\centering
\includegraphics[width=0.65\linewidth]{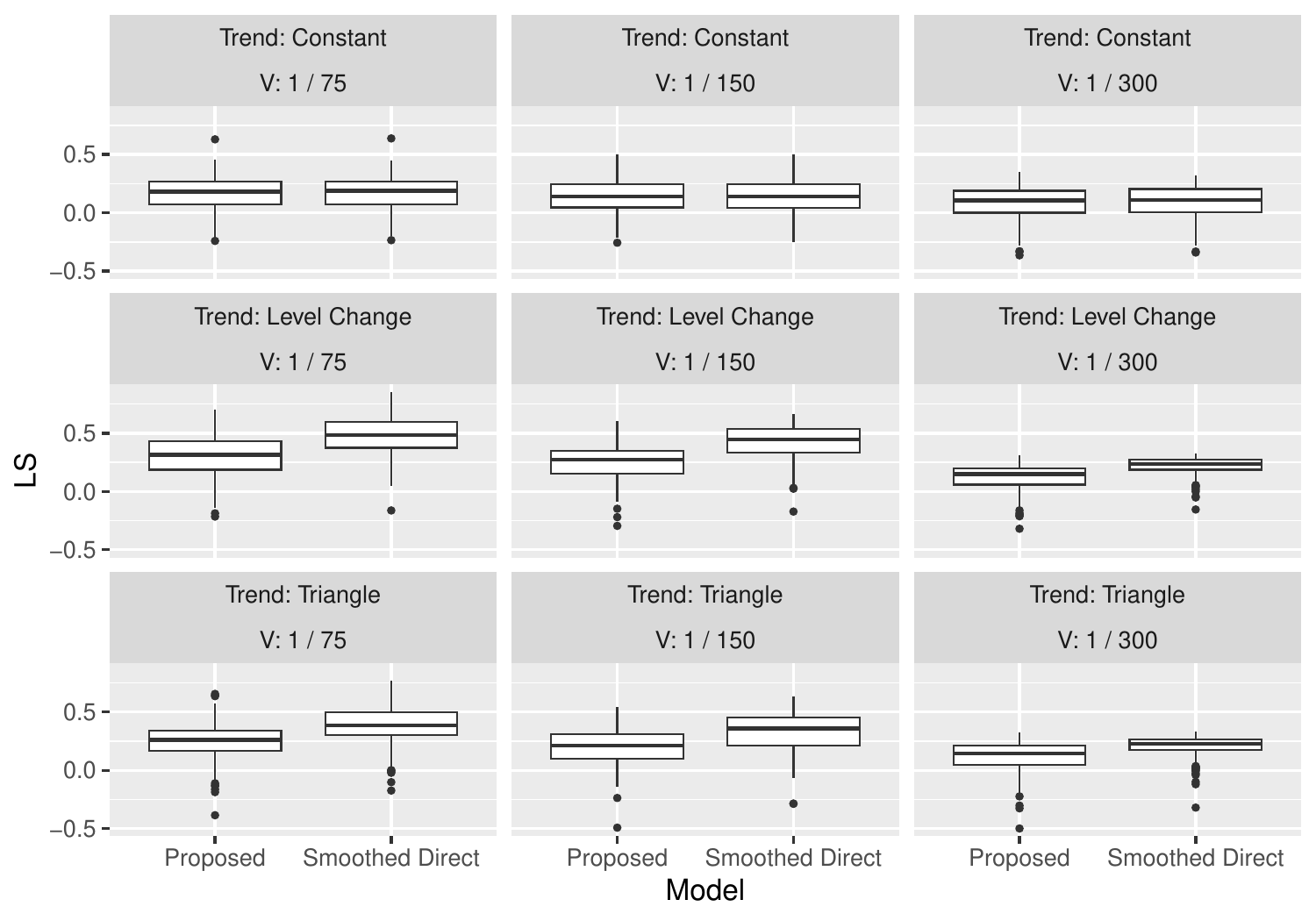}
\caption{LS for the proposed and smoothed direct models for  simulation settings where $\tau_i$ is not the same for all time points.}
\label{fig:cpo_plots_unequal}
\end{figure}

In Figures \ref{fig:rmse_logit_risk_plots_equal} and \ref{fig:rmse_logit_risk_plots_unequal} we plot RMSE for the simulation settings where $\tau_i$ is the same for all time points and where $\tau_i$ is not the same for all time points, respectively. {The models have nearly identical performance under RMSE.} In Figures \ref{fig:dic_plots_equal} and \ref{fig:dic_plots_unequal} we plot DIC for the simulation settings where $\tau_i$ is the same for all time points and where $\tau_i$ is not the same for all time points, respectively. {The proposed model outperforms the smoothed direct model under DIC when $V$ is larger, and the trend is non-constant.} In Figures \ref{fig:cpo_plots_equal} and \ref{fig:cpo_plots_unequal} we plot LS for the simulation settings where $\tau_i$ is the same for all time points and where $\tau_i$ is not the same for all time points, respectively. {The proposed model outperforms the smoothed direct model under LS when the trend is non-constant.}




\section{Rwanda Application Appendix}
In Table \ref{tab:par_ests_rwanda} we summarize of the posterior for the  various parameters in the smoothed direct and proposed models.
\begin{table}[!ht]
\centering
\caption{Comparison of parameter estimates for the smoothed direct and proposed models in the Rwanda application.}
\label{tab:par_ests_rwanda}
\begin{tabular}{llrrrrr}
  \hline
Model & Parameter & Mean & SD & 2.5\% Quantile & Median & 97.5\% Quantile\\ 
  \hline
 Smoothed Direct & $\mu$ & -1.25 & 0.52 & -2.28 & -1.25 & -0.23 \\ 
  Smoothed Direct & $\beta$ & -0.05 & 0.03 & -0.11 & -0.05 & 0.01 \\ 
  Smoothed Direct & $\tau$ & 7.82 & 2.68 & 3.62 & 7.50 & 13.98 \\ 
  Smoothed Direct & $\phi$ & 0.95 & 0.05 & 0.82 & 0.97 & 1.00 \\ 
  Smoothed Direct & $\nu_{1992}$ & 0.02 & 0.04 & -0.05 & 0.02 & 0.11 \\ 
  Smoothed Direct & $\nu_{2000}$ & 0.00 & 0.03 & -0.06 & 0.00 & 0.07 \\ 
  Smoothed Direct & $\nu_{2005}$ & 0.04 & 0.03 & -0.02 & 0.04 & 0.12 \\ 
  Smoothed Direct & $\nu_{2008}$ & -0.07 & 0.04 & -0.16 & -0.07 & 0.00 \\ 
  Smoothed Direct & $\nu_{2010}$ & 0.00 & 0.03 & -0.07 & 0.00 & 0.07 \\ 
  Smoothed Direct & $\nu_{2015}$ & 0.00 & 0.04 & -0.07 & 0.00 & 0.07 \\ 
  \hline
  Proposed & $\mu$ & -0.99 & 0.45 & -1.90 & -0.98 & -0.12 \\ 
  Proposed & $\beta$ & -0.06 & 0.03 & -0.11 & -0.06 & -0.01 \\ 
  Proposed & $\tau$ & 13.80 & 5.58 & 5.43 & 12.99 & 26.98 \\ 
  Proposed & $\phi$ & 0.96 & 0.05 & 0.84 & 0.98 & 1.00 \\ 
  Proposed & $\theta$ & 0.31 & 0.16 & 0.09 & 0.28 & 0.68 \\ 
  Proposed & $\nu_{1992}$ & 0.02 & 0.04 & -0.05 & 0.02 & 0.11 \\ 
  Proposed & $\nu_{2000}$ & 0.00 & 0.03 & -0.06 & 0.00 & 0.07 \\ 
  Proposed & $\nu_{2005}$ & 0.05 & 0.03 & -0.01 & 0.04 & 0.12 \\ 
  Proposed & $\nu_{2008}$ & -0.07 & 0.04 & -0.16 & -0.07 & 0.00 \\ 
  Proposed & $\nu_{2010}$ & 0.00 & 0.03 & -0.07 & 0.00 & 0.07 \\ 
  Proposed & $\nu_{2015}$ & 0.00 & 0.04 & -0.07 & 0.00 & 0.07 \\ 
   \hline
\end{tabular}
\end{table}

\section{Estimation of U5MR across Multiple Countries}
\label{sec:mc_app}
In this appendix we will simultaneously estimate U5MR subnationally at the Admin1 level across multiple countries during the 2010-2014 time period. Specifically, we will consider Burundi, Ethiopia, Kenya, Rwanda, Tanzania, and Uganda using DHS surveys from 2016, 2016, 2014, 2015, 2015, and 2016 respectively. First, to understand between-country variation, we will fit separate smoothed direct models to each country, with ICARs for the structured spatial random effect. We will then consider two sets of models to compare, where each model is fit simultaneously to the six target countries:
\begin{enumerate}
    \item The smoothed direct model with an ICAR for the structured spatial random effect, which we will refer to as the \textit{smoothed direct model}, and the smoothed direct model with our proposed multi-country AICAR for the structured spatial random effect, which we will refer to as the \textit{proposed model}.
    \item The smoothed direct model and proposed model, but replacing the intercept $\mu$ in the smoothed direct model described in
    Section 2.2 of the main text with country-specific intercepts $\mu_{c[i]}$, where $c[i]$ denotes the country in which region $i$ resides. We will refer to these models as the \textit{smoothed direct country-intercept model} and \textit{proposed country-intercept model}.
\end{enumerate}

\subsection{Country-Specific Models}
In Table \ref{tab:par_ests_country-specific} we display summaries of the posterior for the various parameters from fitting the smoothed direct model to each country separately. The posterior summaries for the intercepts for Kenya and Rwanda are smaller than intercepts for the rest of the countries, indicating that U5MR is on average lower in these countries across Admin1 regions. All countries except Burundi have roughly comparable posterior summaries for $\phi$: a posterior mode close to 0 and a posterior median close to 0.3. This indicates for these countries that there is not much weight being placed on the structured spatial random effect. Burundi has a posterior mode at 0.23 and a posterior median of 0.43, so while again there is not much weight being placed on the structured spatial random effect, there is still more spatial smoothing occurring than in the other five countries. The parameter with the most heterogeneity across countries is $\tau$. Besides Uganda, the countries have posterior medians from around 10-30, with varying posterior standard deviations from around 5 to 70. Uganda however has a very large posterior summaries for precision, with posterior median of 457. This indicates that there is a large amount of heterogeneity in how the smoothed direct models for each country weight the total region-level random effect. Uganda in particular has such large posterior summaries for $\tau$ that the Admin1 region estimates are heavily smoothed towards the country's direct estimate.

\begin{table}[!ht]
\centering
\caption{Comparison of parameter estimates from fitting the smoothed direct model to each country separately.}
\label{tab:par_ests_country-specific}
\begin{tabular}{llrrrrrr}
  \hline
Country & Parameter & Mean & SD & 2.5\% Quantile & Median & 97.5\% Quantile & Mode \\ 
  \hline
Burundi & $\mu$ & -2.60 & 0.07 & -2.74 & -2.60 & -2.46 & -2.60 \\ 
  Burundi & $\tau$ & 11.58 & 5.36 & 4.37 & 10.53 & 24.96 & 8.69 \\ 
  Burundi & $\phi$ & 0.45 & 0.25 & 0.06 & 0.43 & 0.91 & 0.23 \\ 
  \hline
  Ethiopia & $\mu$ & -2.49 & 0.08 & -2.65 & -2.48 & -2.33 & -2.48 \\ 
  Ethiopia & $\tau$ & 21.66 & 15.80 & 5.06 & 17.44 & 63.38 & 11.56 \\ 
  Ethiopia & $\phi$ & 0.31 & 0.25 & 0.02 & 0.24 & 0.86 & 0.03 \\ 
  \hline
  Kenya & $\mu$ & -2.94 & 0.07 & -3.09 & -2.95 & -2.79 & -2.95 \\ 
  Kenya & $\tau$ & 47.57 & 47.91 & 7.16 & 33.45 & 173.17 & 17.61 \\ 
  Kenya & $\phi$ & 0.31 & 0.25 & 0.01 & 0.24 & 0.87 & 0.03 \\ 
  \hline
  Rwanda & $\mu$ & -3.04 & 0.11 & -3.28 & -3.04 & -2.83 & -3.03 \\ 
  Rwanda & $\tau$ & 44.53 & 70.11 & 3.64 & 24.17 & 211.64 & 9.09 \\ 
  Rwanda & $\phi$ & 0.35 & 0.26 & 0.02 & 0.29 & 0.91 & 0.04 \\ 
  \hline
  Tanzania & $\mu$ & -2.63 & 0.07 & -2.77 & -2.63 & -2.50 & -2.63 \\ 
  Tanzania & $\tau$ & 18.64 & 12.02 & 5.63 & 15.46 & 50.22 & 11.17 \\ 
  Tanzania & $\phi$ & 0.32 & 0.26 & 0.01 & 0.25 & 0.90 & 0.02 \\ 
  \hline
  Uganda & $\mu$ & -2.61 & 0.05 & -2.71 & -2.61 & -2.52 & -2.61 \\ 
  Uganda & $\tau$ & 3783.69 & 37696.07 & 18.76 & 457.11 & 25111.91 & 34.55 \\ 
  Uganda & $\phi$ & 0.37 & 0.27 & 0.02 & 0.31 & 0.91 & 0.05 \\ 
   \hline
\end{tabular}
\end{table}

\subsection{Smoothed Direct and Proposed Models}

In Table \ref{tab:par_ests_no_country-int} we display summaries of the posterior for the  various parameters from fitting the smoothed direct and proposed models. We see that the posterior summaries for $\mu$ is comparable between models. However, the posterior summaries for $\tau$ and $\phi$ are all larger under the proposed model. In particular, the larger posterior summaries of $\phi$ in the proposed model indicate that it is placing more weight on the structured spatial random effect.

\begin{table}[!ht]
\centering
\caption{Comparison of parameter estimates for the smoothed direct and proposed models in the multi-country application.}
\label{tab:par_ests_no_country-int}
\begin{tabular}{llrrrrrr}
  \hline
Model & Parameter & Mean & SD &  2.5\% Quantile & Median & 97.5\% Quantile & Mode \\ 
  \hline
Smoothed Direct & $\mu$ & -2.67 & 0.04 & -2.75 & -2.67 & -2.59 & -2.67 \\ 
  Smoothed Direct & $\tau$ & 10.23 & 2.39 & 6.33 & 9.96 & 15.67 & 9.45 \\ 
  Smoothed Direct & $\phi$ & 0.23 & 0.22 & 0.01 & 0.15 & 0.80 & 0.01 \\ 
  \hline
  Proposed & $\mu$ & -2.67 & 0.03 & -2.74 & -2.67 & -2.60 & -2.67 \\ 
  Proposed & $\tau$ & 12.42 & 3.48 & 6.89 & 11.99 & 20.46 & 11.17 \\ 
  Proposed & $\phi$ & 0.49 & 0.23 & 0.09 & 0.49 & 0.90 & 0.49 \\ 
  Proposed & $\theta$ & 0.24 & 0.16 & 0.04 & 0.19 & 0.65 & 0.11 \\ 
   \hline
\end{tabular}
\end{table}

\begin{figure}[!ht]
\centering
\includegraphics[width=0.99\linewidth]{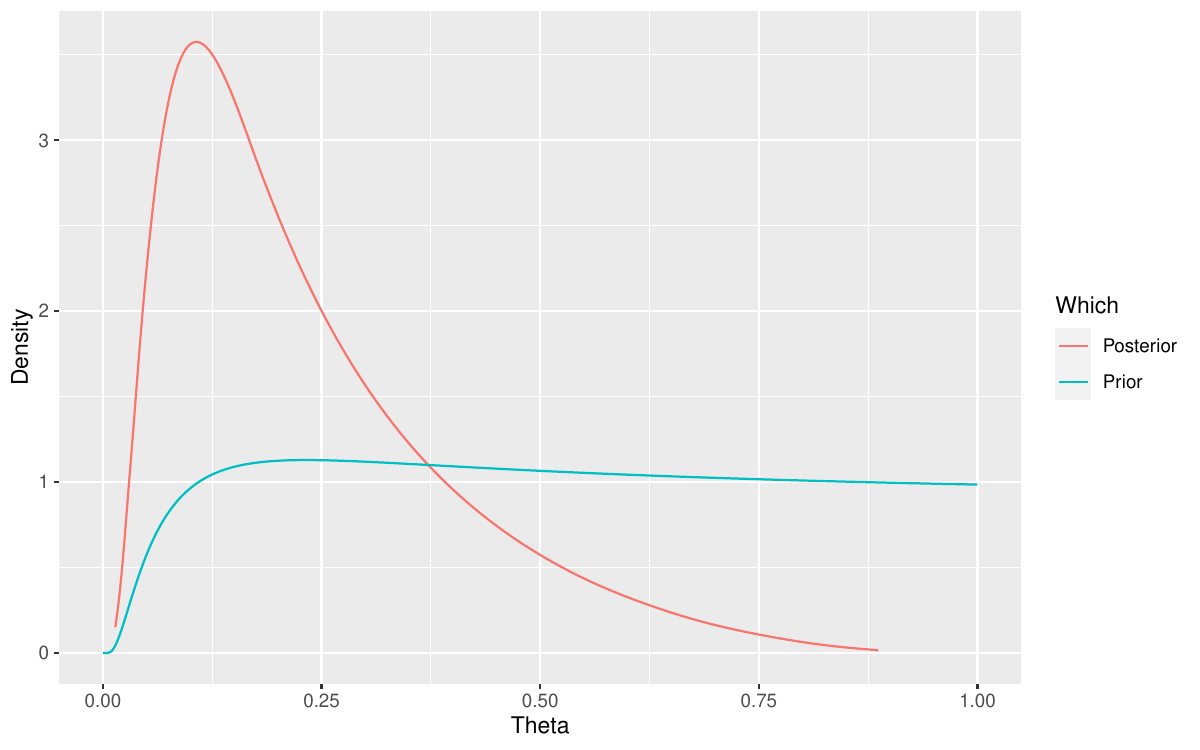}
\caption{Comparison of prior and posterior density for $\theta$ in the proposed model for the multi-country application.}
\label{fig:theta_comp_no_country-int}
\end{figure}

Focusing on $\theta$, in Figure \ref{fig:theta_comp_no_country-int} we plot the prior and posterior for $\theta$ from the proposed model. We see that the prior is relatively flat from 0.2 to 1, with a mode at 0.23, and $95\%$ of the prior mass lying in $[0.07, 0.97]$. The posterior has a mode at 0.11, which it is heavily concentrated around, and the $95\%$ credible interval for $\theta$ is $[0.04, 0.65]$. Combining this with the posterior summaries for $\phi$, we find that the proposed model is performing more spatial smoothing than the smoothed direct model, as $\phi$ is larger in the proposed model, but there is not much spatial smoothing occurring on the borders of different countries, as $\theta$ is close to 0 in the proposed model.

\begin{figure}[!ht]
\centering
\includegraphics[width=0.99\linewidth]{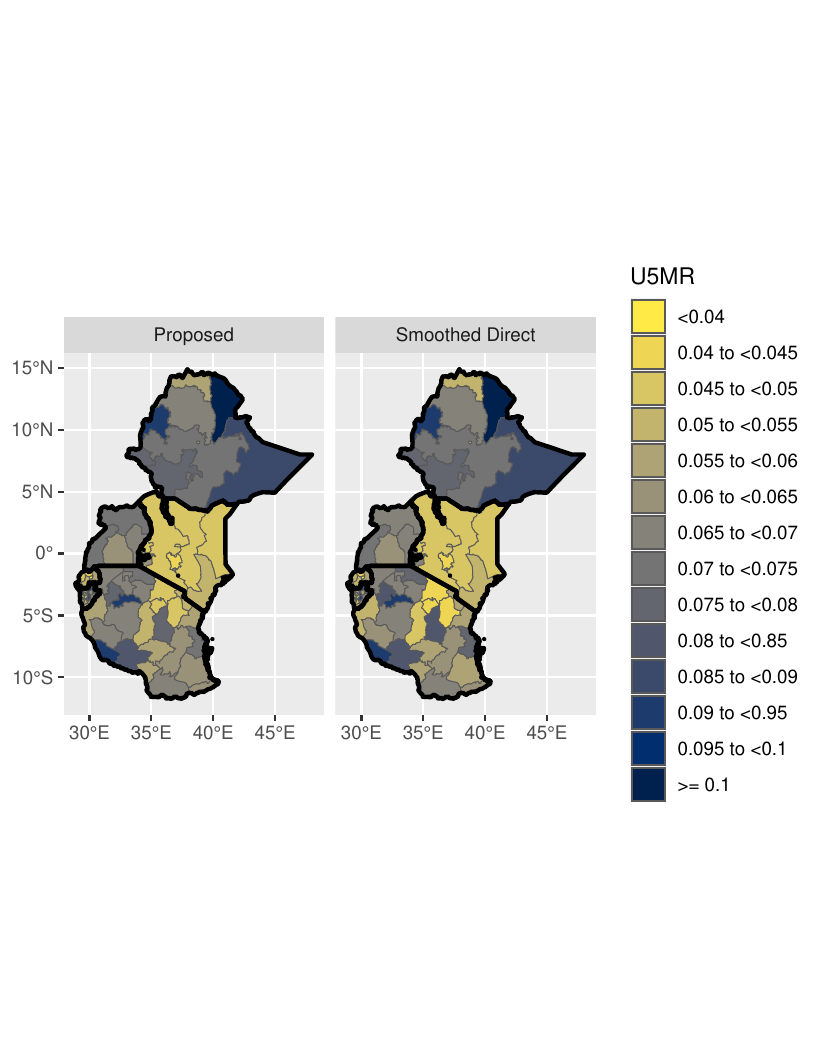}
\caption{Maps of U5MR estimates from the smoothed direct model and the proposed model.}
\label{fig:mc_estimates_no_country-int_1}
\end{figure}

\begin{figure}[!ht]
\centering
\includegraphics[width=1.1\linewidth]{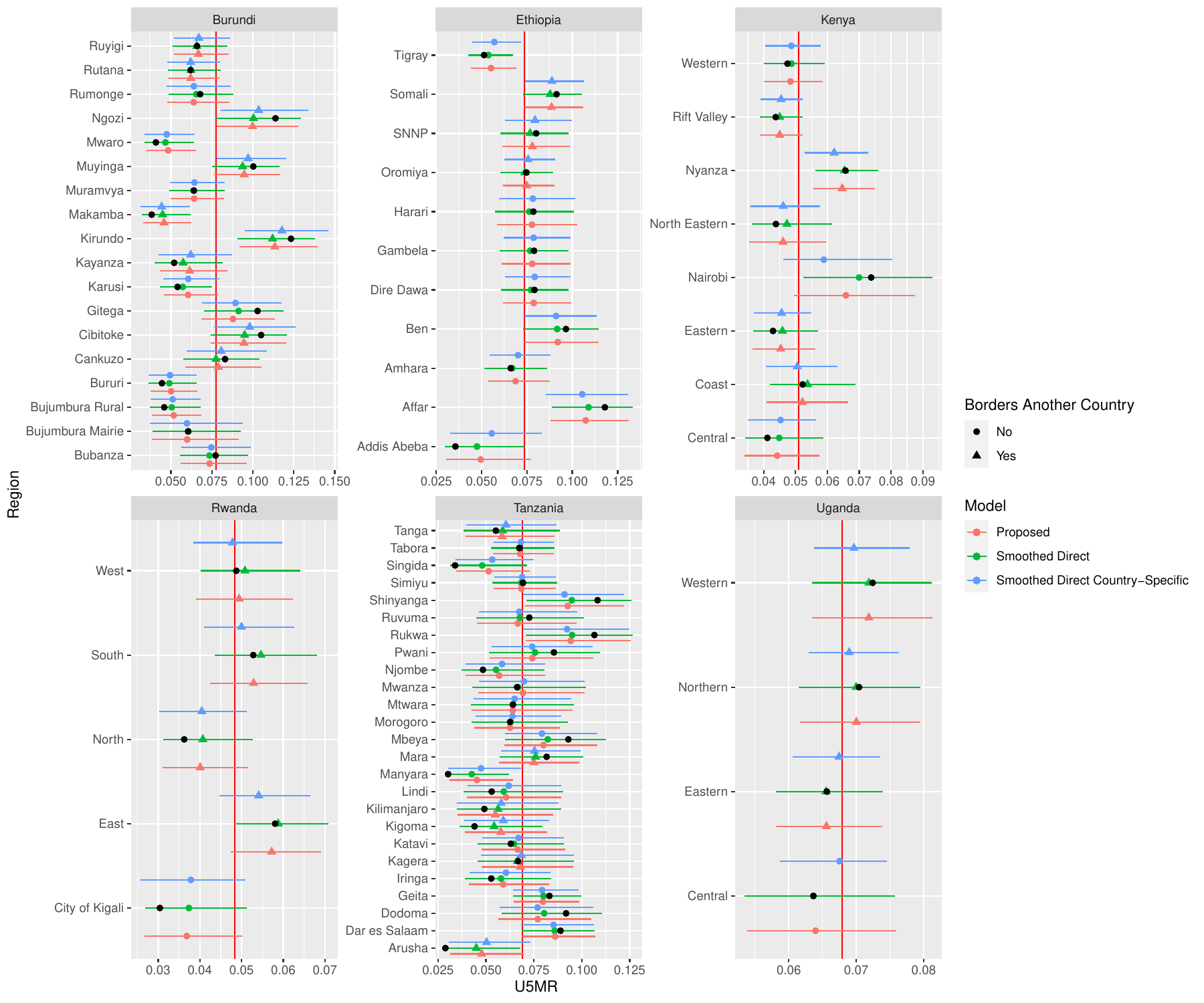}
\caption{Comparison of U5MR estimates from the smoothed direct model and the proposed model, in addition to the country-specific smoothed direct model fits, direct estimates for each Admin1 region in black, and direct estimates for each country in red.}
\label{fig:mc_estimates_no_country-int_2}
\end{figure}

In Figure \ref{fig:mc_estimates_no_country-int_1} we map U5MR estimates from the smoothed direct model and the proposed model. To get a better grasp on how the estimates differ between the two models, in Figure \ref{fig:mc_estimates_no_country-int_2} we plot U5MR estimates from the smoothed direct model and the proposed model, in addition to the country-specific smoothed direct model fits, direct estimates for each Admin1 region, and direct estimates for each country. We also include whether each Admin1 region borders a different country. The estimates from the smoothed direct and proposed models differ the most among Admin1 regions that do not border a different country. In particular, in these regions, the estimates from the proposed model are pulled towards their country level direct estimate. This can be explained by the proposed model estimating larger values of $\phi$. This places more weight on the structured spatial random effect, leading to more spatial smoothing within countries.

\subsection{Smoothed Direct and Proposed Country-Intercept Models}
In Table \ref{tab:par_ests_country-int} we display summaries of the posterior for the  various parameters from fitting the smoothed direct and proposed country-intercept models. We see that the posterior point estimates for the country-specific intercepts are comparable between models, whereas the posterior standard deviations are larger in the proposed country-intercept model. The posterior summaries are roughly comparable for $\tau$ and $\phi$, with the smoothed direct country-intercept model having slightly larger summaries for $\phi$ and the proposed country-intercept model having slightly larger summaries for $\tau$.

\begin{table}[!ht]
\centering
\caption{Comparison of parameter estimates for the smoothed direct and proposed country-intercept models in the multi-country application.}
\label{tab:par_ests_country-int}
\begin{tabular}{lllrrrrrr}
  \hline
Model & Country & Param. & Mean & SD & 2.5\% Quantile & Median & 97.5\% Quantile & Mode \\ 
  \hline
Smoothed Direct C-Int. & Burundi & $\mu$ & -2.56 & 0.13 & -2.83 & -2.56 & -2.29 & -2.57 \\ 
  Smoothed Direct C-Int. & Ethiopia & $\mu$ & -2.36 & 0.21 & -2.75 & -2.37 & -1.91 & -2.41 \\ 
  Smoothed Direct C-Int. & Kenya & $\mu$ & -2.93 & 0.14 & -3.21 & -2.93 & -2.65 & -2.93 \\ 
  Smoothed Direct C-Int. & Rwanda & $\mu$ & -3.17 & 0.18 & -3.54 & -3.16 & -2.82 & -3.16 \\ 
  Smoothed Direct C-Int. & Tanzania & $\mu$ & -2.69 & 0.11 & -2.92 & -2.68 & -2.49 & -2.67 \\ 
  Smoothed Direct C-Int. & Uganda & $\mu$ & -2.67 & 0.18 & -3.03 & -2.67 & -2.32 & -2.66 \\ 
  Smoothed Direct C-Int. &  & $\tau$ & 12.80 & 3.96 & 6.78 & 12.21 & 22.19 & 11.12 \\ 
  Smoothed Direct C-Int. &  & $\phi$ & 0.57 & 0.26 & 0.08 & 0.60 & 0.96 & 0.86 \\ 
  \hline
  Proposed C-Int. & Burundi & $\mu$ & -2.57 & 0.18 & -2.94 & -2.57 & -2.20 & -2.57 \\ 
  Proposed C-Int. & Ethiopia & $\mu$ & -2.36 & 0.30 & -2.94 & -2.38 & -1.73 & -2.41 \\ 
  Proposed C-Int. & Kenya & $\mu$ & -2.93 & 0.19 & -3.31 & -2.93 & -2.54 & -2.93 \\ 
  Proposed C-Int. & Rwanda & $\mu$ & -3.17 & 0.24 & -3.68 & -3.17 & -2.71 & -3.15 \\ 
  Proposed C-Int. & Tanzania & $\mu$ & -2.68 & 0.14 & -2.97 & -2.68 & -2.42 & -2.67 \\ 
  Proposed C-Int. & Uganda & $\mu$ & -2.67 & 0.26 & -3.20 & -2.67 & -2.15 & -2.66 \\ 
  Proposed C-Int. &  & $\tau$ & 13.49 & 3.97 & 7.38 & 12.92 & 22.86 & 11.84 \\ 
  Proposed C-Int. &  & $\phi$ & 0.53 & 0.25 & 0.08 & 0.54 & 0.94 & 0.71 \\ 
  Proposed C-Int. &  & $\theta$ & 0.51 & 0.27 & 0.06 & 0.50 & 0.95 & 0.29 \\ 
   \hline
\end{tabular}
\end{table}

\begin{figure}[!ht]
\centering
\includegraphics[width=0.99\linewidth]{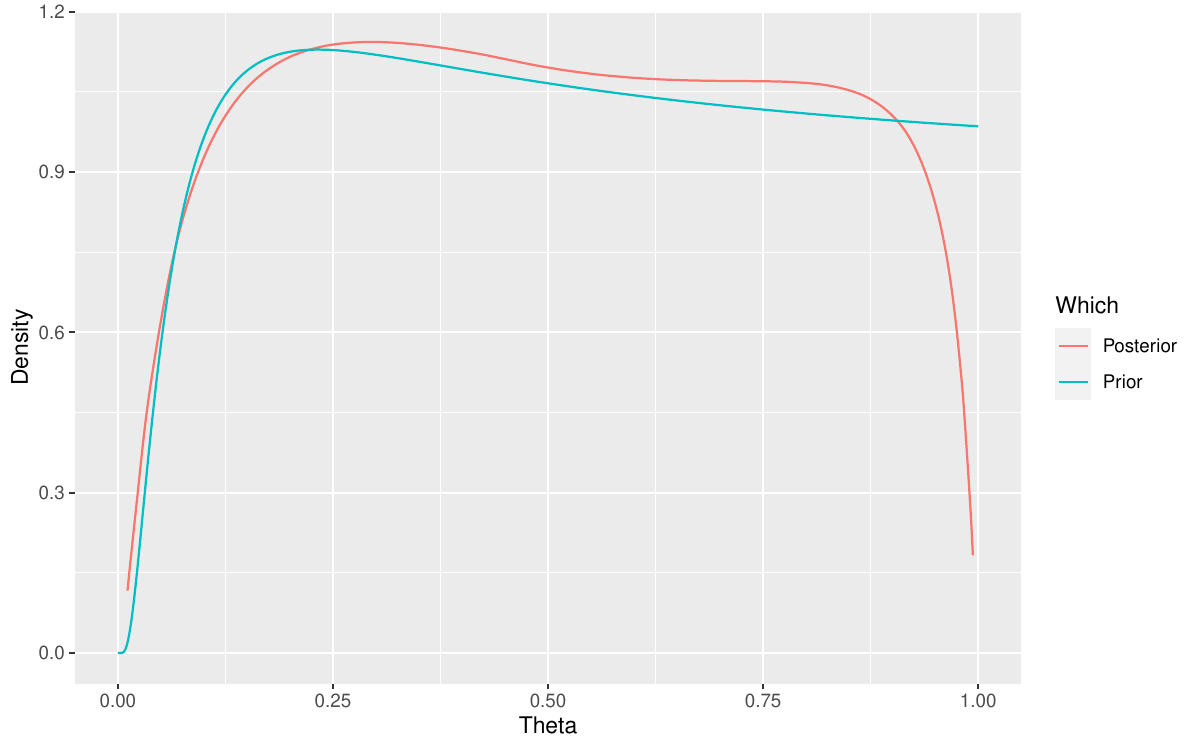}
\caption{Comparison of prior and posterior density for $\theta$ in the proposed country-intercept model for the multi-country application.}
\label{fig:theta_comp_country-int}
\end{figure}

Focusing on $\theta$, in Figure \ref{fig:theta_comp_country-int} we plot the prior and posterior for $\theta$ from the proposed model. The posterior has a mode at 0.29, which it is not heavily concentrated around, and the $95\%$ credible interval for $\theta$ is $[0.06, 0.95]$. We see that the posterior is very close to the prior, with the main difference being that the posterior does not place much mass immediately around 1. Thus the data did not inform the posterior for $\theta$ beyond the fact that it is not 1.

\begin{figure}[!ht]
\centering
\includegraphics[width=0.99\linewidth]{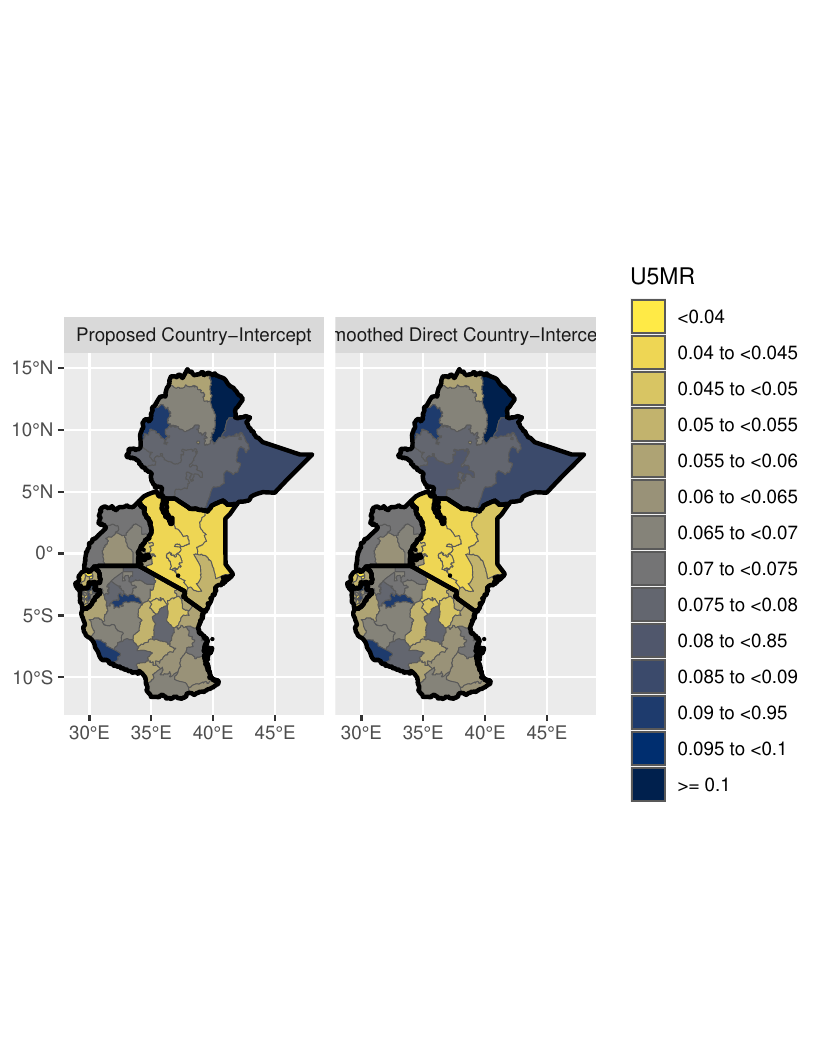}
\caption{Maps of U5MR estimates from the smoothed direct and proposed country-intercept models.}
\label{fig:mc_estimates_country-int_1}
\end{figure}

\begin{figure}[!ht]
\centering
\includegraphics[width=1.1\linewidth]{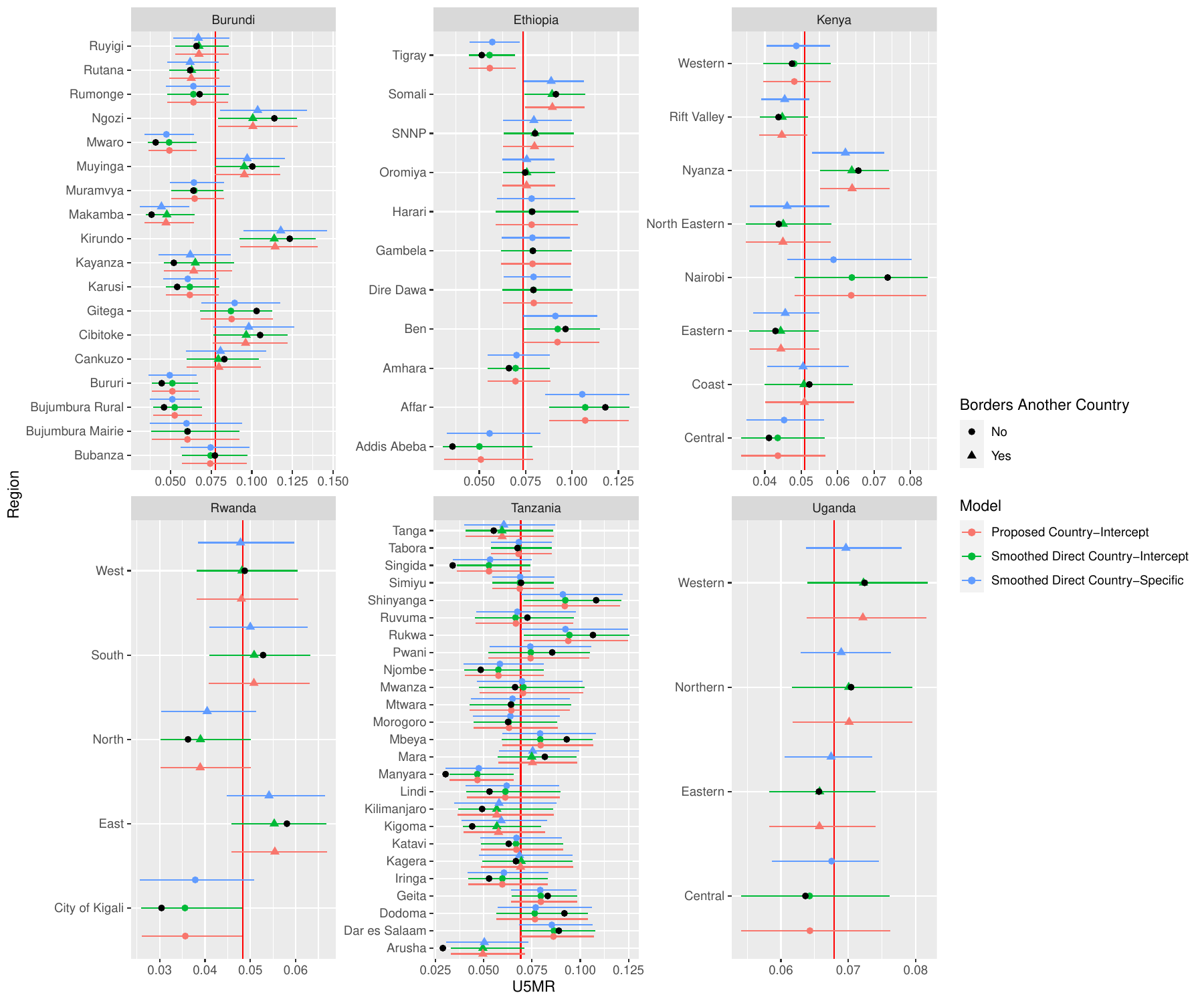}
\caption{Comparison of U5MR estimates from the smoothed direct and proposed country-intercept models, in addition to the country-specific smoothed direct model fits, direct estimates for each Admin1 region in black, and direct estimates for each country in red.}
\label{fig:mc_estimates_country-int_2}
\end{figure}

In Figure \ref{fig:mc_estimates_country-int_1} we map U5MR estimates from the smoothed direct and proposed country-intercept models. To get a better grasp on how the estimates differ between the two models, in Figure \ref{fig:mc_estimates_country-int_2} we plot U5MR estimates from the smoothed direct and proposed country-intercept models, in addition to the country-specific smoothed direct model fits, direct estimates for each Admin1 region, and direct estimates for each country. We also include whether each Admin1 region borders a different country. The smoothed direct and proposed country-intercept models have close to identical estimates in nearly all regions.

In this multi-country application we have found that while our AICAR improves estimate of U5MR when country-specific intercepts are not used, it does not provide benefit for estimation of U5MR in the presence of country-specific intercepts. Country-specific intercepts are not always used in such mutli-country settings \citep[see e.g.][]{Burstein_2019}, but we believe that they are natural to consider. In contrast, in the temporal Rwanda application in Section 6 of the main text, the results did not substantively change when conflict-period specific intercepts were introduced. Thus, it would seem that the benefit of our AGMRFs is application dependent. We conjecture that our AICAR would improve estimation of U5MR, when country-specific intercepts are used, in settings where some of the countries have much less data than other countries. This data scarcity would make the structured spatial random effect more influential, and it would be interesting to see in such a setting whether the AICAR would improve over the performance of an ICAR for the structured spatial random effect. In Appendix \ref{sec:general_mc}, we consider a more general multi-country AICAR.

\section{A More General Multi-Country Model}
\label{sec:general_mc}
In the multi-country application in Appendix \ref{sec:mc_app}, we found that our AICAR provided little benefit over including country-specific intercepts in the model. In this appendix we present a more general version of the adaptive ICAR for multiple countries, to determine if a more flexible random effect specification will provide improved performance in the multi-country application.

\subsection{Model Description}
Suppose we are specifying a structured spatial random effect for $N$ regions at the Admin1 level with no islands, which are nested within $M$ countries. For neighboring Admin1 regions $i\sim j$, let $A_{ij}$ be an indicator that $i$ and $j$ are nested within the same country. For Admin1 region $i$, we let $c[i]$ denote the country region $i$ lies in. Suppose for neighboring Admin1 regions $i\sim j$ we specify
\begin{align*}
    x_i-x_j\mid \tau_{1}, \dots, \tau_{M+1}  &\sim\textsf{Normal}(0,\tau_{c[i]}^{-1}) & \text{ if } A_{ij}=1, \\
    x_i-x_j\mid \tau_{1}, \dots, \tau_{M+1}  &\sim\textsf{Normal}(0,\tau_{M+1}^{-1}) & \text{ if } A_{ij}=0.
\end{align*}
This is a simplified AICAR with $M+1$ precisions: $M$ precisions for neighboring Admin1 regions within the same country, $\tau_1, \dots, \tau_M$, and one precision for neighboring Admin1 regions between different countries, $\tau_{M+1}$. We note that for the multi-country application, $M=6$ is small enough that we can fit this model in \texttt{INLA}. With a larger number of countries $M$ this may not be feasible.

As discussed in Section 2.3.2 of the main text, we expect that $\tau_m>\tau_{M+1}$ for each $m\in[M]$. It follows that $\bx\mid \tau_{1}, \dots, \tau_{M+1} \sim \textsf{Normal}(0,Q^-)$, where $Q$ is a special case of the general AICAR precision matrix.  We can simplify $Q$ to $Q = \sum_{m=1}^{M+1}\tau_m R_m$, where for $m\in[M+1]$, $R_m=D_m-W_m$ where
\[
W_{m,{ij}}=
    \begin{cases}
      I(i\sim j, A_{ij}=1,  \text{ and } c[i]=m), & \text{if}\ m< M + 1 \\
      I(i\sim j \text{ and } A_{ij}=0), & \text{if}\ m = M+1,
    \end{cases}
\]
and
$D_m=\text{diag}\left(\sum_{j=1}^{N}W_{m, {1j}},\dots, \sum_{j=1}^{N}W_{m, {Nj}}\right)$. Note that when $\tau_1 =\dots=\tau_{M+1}$ this random effect specification reduces to an ICAR as presented in Section 2.1.2 of the main text. In particular we have in this case that  $Q = \tau_1( \sum_{m=1}^{M+1} R_m)$ where $\sum_{m=1}^{M+1}R_m$ is the structure matrix of an ICAR.

\subsection{Scaling, Reparameterizations, and Prior Choice}
Similar to Section 4 of the main text, we briefly describe various considerations for fitting this more general multi-country model.

\subsubsection{Scaling} Let $\sigma^2(\bx)$ denote the geometric mean of the marginal variances of the elements of $\bx$ when setting $\tau_1 =\dots=\tau_{M+1}=1$ (i.e. so that $Q=\sum_{m=1}^{M+1}R_m$ is the structure matrix of an ICAR). We will work with the following scaled precision matrix $Q^* = \sum_{m=1}^{M+1}\tau_m R_m^*,$
where for $m\in[M+1]$, $R_m^*=R_m/\sigma^2(\bx)$. It then follows that when $\tau_1 =\dots=\tau_{M+1}$, $1/\tau_1$ represents the approximate marginal variance of $\bx$, independent of the structure matrix $\sum_{m=1}^{M+1}R_m$.

\subsubsection{Interpretability} We will reparameterize the precision matrix as
\[Q^* = \sum_{m=1}^{M+1}\tau_m R_m^*=\tau_1\left(\sum_{m=1}^{M+1}\tau_m/\tau_1 R_m^*\right)=\tau_1\left(R_1^*+\sum_{m=2}^{M}\psi_m R_m^* + \theta R_{M+1}^*\right),\]
so that $\psi_m=\tau_m/\tau_1\in[0,\infty)$ for $m=2,\dots, M$, and $\theta=\tau_{M+1}/\tau_1\in[0,1]$. With this parameterization, $\tau_1$ and $\theta$ can be interpreted as in the AICAR described in Section 4 of the main text, just with respect to the first country (which will be Burundi in the multi-country application). The $\psi_m$ represents the ratio of country $m$'s precision to the first country's precision.

\subsubsection{BYM2-Like Parameterization} Using the reparameterization just introduced, a BYM2-like paramterization follows just as in Section 4 of the main text, with $R_1^*+\theta R_2^*$ replaced by $R_1^*+\sum_{m=2}^{M}\psi_m R_m^* + \theta R_{M+1}^*$.

\subsubsection{Prior Specification} For simplicity we will use the same priors as in Section 4.3 of the main text for all applicable parameters. We will use the PC prior for precisions for each $\psi_m$, i.e. the prior described in Section 4.3.1 of the main text.

\subsection{Results}

We fit the smoothed direct model with our more general multi-country AICAR for the structured spatial random effect, which we will refer to as the \textit{proposed general model}, to the multi-country application from Appendix \ref{sec:mc_app}. In Table \ref{tab:par_ests_general} we display summaries of the posterior for the various parameters from fitting the proposed general model. We see that the estimates for $\mu$ and $\theta$ are comparable to those from the proposed model in Table \ref{tab:par_ests_no_country-int}. However, $\phi$ has slightly larger posterior summaries than in the proposed model, indicating there is slightly more spatial smoothing occurring in the proposed general model compared to the proposed model. The posterior for $\tau$ for Burundi, the reference country, is roughly similar to the posterior for $\tau$ in the proposed model, even though it has a different interpretation. The $psi$ parameters for the remaining countries are all positive, except for Rwanda.

\begin{table}[ht]
\centering
\caption{Parameter estimates for the proposed general model in the multi-country application.}
\label{tab:par_ests_general}
\begin{tabular}{lllrrrrrr}
  \hline
Model & Country & Param. & Mean & SD & 2.5\% Quantile & Median & 97.5\% Quantile & Mode \\ 
  \hline
Proposed General &  & $\mu$ & -2.67 & 0.04 & -2.74 & -2.67 & -2.60 & -2.67 \\ 
  Proposed General &  & $\phi$ & 0.56 & 0.23 & 0.12 & 0.58 & 0.93 & 0.71 \\ 
  Proposed General &  & $\theta$ & 0.25 & 0.17 & 0.04 & 0.21 & 0.66 & 0.12 \\ 
  Proposed General & Burundi & $\tau$ & 12.22 & 4.42 & 5.62 & 11.53 & 22.80 & 10.23 \\ 
  Proposed General & Ethiopia & $\psi$ & 5.22 & 12.23 & 0.09 & 1.73 & 32.61 & 0.35 \\ 
  Proposed General & Kenya & $\psi$ & 21.73 & 76.25 & -0.08 & 4.26 & 162.92 & 0.36 \\ 
  Proposed General & Rwanda & $\psi$ & 0.91 & 1.77 & 0.02 & 0.36 & 5.26 & 0.06 \\ 
  Proposed General & Tanzania & $\psi$ & 1.68 & 2.33 & 0.14 & 0.93 & 7.83 & 0.36 \\ 
  Proposed General & Uganda & $\psi$ & 20.22 & 73.89 & -0.38 & 3.42 & 153.28 & 0.13 \\ 
   \hline
\end{tabular}
\end{table}

\begin{figure}[!ht]
\centering
\includegraphics[width=0.99\linewidth]{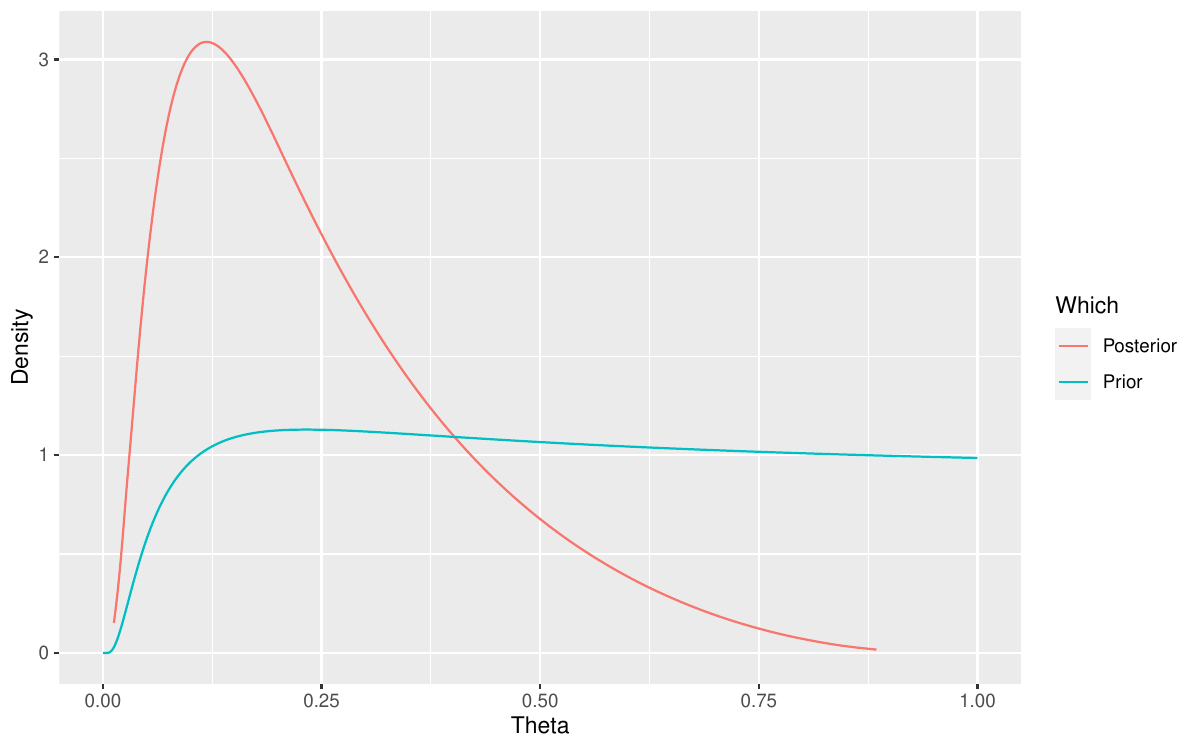}
\caption{Comparison of prior and posterior density for $\theta$ in the proposed general model for the multi-country application.}
\label{fig:theta_comp_general}
\end{figure}

\begin{figure}[!ht]
\centering
\includegraphics[width=1.1\linewidth]{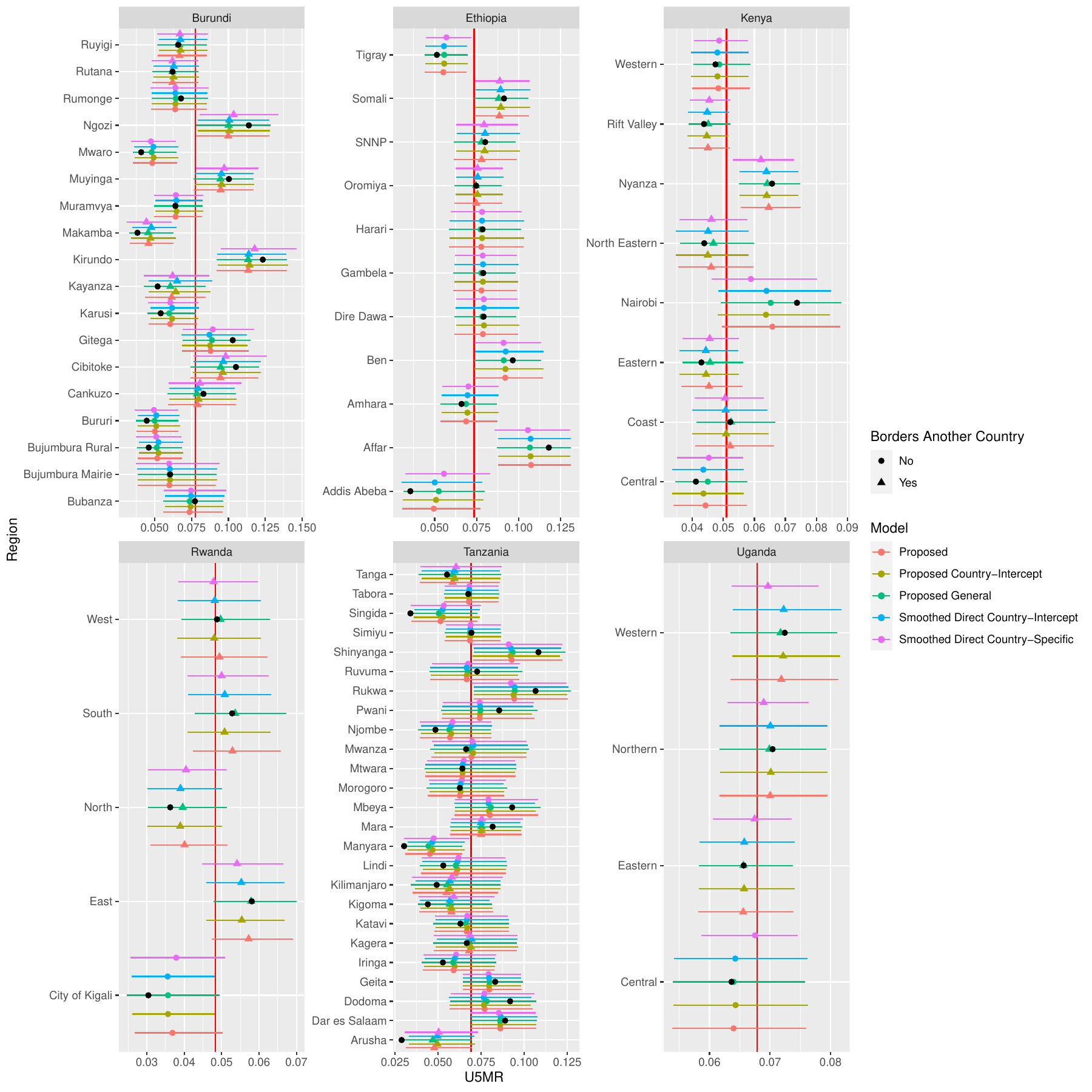}
\caption{Comparison of U5MR estimates from the proposed general model to those from the proposed, proposed country-intercept, and smoothed direct country-intercept models, in addition to the country-specific smoothed direct model fits, direct estimates for each Admin1 region in black, and direct estimates for each country in red.}
\label{fig:mc_estimates_country-int_2_general}
\end{figure}

In Figure \ref{fig:theta_comp_general} we plot the prior and posterior for $\theta$ from the proposed general model. In Figure \ref{fig:mc_estimates_country-int_2_general} we plot U5MR estimates from the proposed general model to compare to those from the proposed, proposed country-intercept, and smoothed direct country-intercept models, in addition to the country-specific smoothed direct model fits, direct estimates for each Admin1 region, and direct estimates for each country. We also include whether each Admin1 region borders a different country. Overall, we find that the results from the proposed general model are very similar to the proposed model, and have no evidence that the more flexible random effect specification improved the performance of the proposed model in the multi-country application.

\bibliographystyle{biorefs}
\bibliography{refs}